\documentclass{IEEEtran}
\usepackage{cite}
\usepackage[latin1]{inputenc}
\usepackage{amsmath,amssymb,amsfonts}
\usepackage{algorithmic}
\usepackage{graphicx}
\usepackage{xcolor}
\usepackage{subfig}
\usepackage{diagbox}
\usepackage{ulem}
\usepackage{xcolor}
\usepackage{pdfpages}
\usepackage{multirow}

\DeclareMathOperator{\sinc}{sinc}

\newcommand\sm[1]{\textcolor{black}{#1}}
\newcommand\fs[1]{\textcolor{black}{#1}}
\newcommand\er[1]{\textcolor{black}{#1}}
\newcommand\pb[1]{\textcolor{black}{#1}}
\newcommand\cc[1]{\textcolor{black}{#1}}

\def\BibTeX{{\rm B\kern-.05em{\sc i\kern-.025em b}\kern-.08em
    T\kern-.1667em\lower.7ex\hbox{E}\kern-.125emX}}

\begin{document}

\title{Advanced Analysis of Radar Cross-Section Measurements in Reverberation Environment}
\author{C. Charlo, S. M\'{e}ric, F. Sarrazin,  E. Richalot, J. Sol and P. Besnier,   \IEEEmembership{Senior Member, IEEE}
\thanks{This work is supported by AID/DGA, France. It is also supported in part by the European Union through the European Regional Development Fund, in part by the Ministry of Higher Education and Research, in part by the Région Bretagne, and in part by the Département d'Ille et Vilaine through the CPER Project SOPHIE/STIC \& Ondes.}
\thanks{Corentin Charlo, Stéphane Méric, J\'{e}r\^{o}me Sol and Philippe Besnier are with Univ Rennes, INSA Rennes, CNRS, IETR-UMR 6164, F-35000 Rennes, France (e-mail: \{ccharlo, smeric, jsol, pbesnier\}@insa-rennes.fr}
\thanks{Fran\c{c}ois Sarrazin is with Univ Rennes, CNRS, IETR-UMR 6164, F-35000 Rennes, France (e-mail:francois.sarrazin@univ-rennes.fr).}
\thanks{Elodie Richalot is with Univ Gustave Eiffel, CNRS, ESYCOM, F-77454 Marne-la-Vall\'{e}e, France (e-mail:
elodie.richalot-taisne@univ-eiffel.fr).}
}

\markboth{This work has been submitted to the IEEE for possible publication. Copyright may be transferred without notice.}%
{}

\maketitle

\begin{abstract}
 \fs{Reverberation chambers (RCs) were recently reported as a low-cost alternative to anechoic chambers (ACs) to perform radar cross-section (RCS) pattern measurements}. The method consists of using transmitting and receiving antennas pointing towards a target under test placed on a rotating mast. As a classical RCS characterization, the echo signal is analysed based on two measurements with and without the target in the RC. In the hypothesis of an ideal diffuse field generated in the RC, this signal difference  appears as the echo signal hidden in a Gaussian noise. In case of a point-like backscattering target, observing this signal over a given frequency bandwidth allows the identification of the target response as a sinusoidal signal over this bandwidth whose period is related to the antenna-target distance measured from the measurement calibration plane positions. Therefore, the extraction of the magnitude of this sinusoidal signal requires a proper estimation of this distance. Furthermore, a sinusoidal regression processing relies on the approximation of a constant envelope over the selected frequency bandwidth, imposing some restrictions. In this paper, we introduce a two-step method that consists in identifying the most appropriate distance according to the target's orientation before estimating the magnitude of the sinusoidal signal. We highlight improvement of RCS estimation on a point-like back-scattering target compared to the one-step procedure applied so far. In addition, it is shown that the analysis performed regarding the estimated distance provides a physical insight into the position of the equivalent backscattering point.
\end{abstract}


\section{Introduction}
\fs{\IEEEPARstart{R}{adar} cross-section (RCS) is a well-known physical parameter that quantifies the scattering behavior of an object when illuminated by an electromagnetic wave. Its estimation is of high interest for military applications (e.g. stealth detection) and civil applications (e.g. automotive or air traffic control radars). RCS measurement requires the use of anechoic chambers (ACs) in order to retrieve the line-of-sight contribution between the measurement antenna and the target while minimizing the contributions of all the other propagation paths. However, such measurement environments are costly and not widely available outside some academic or military research laboratories. Therefore, alternative approaches are of interest in order to facilitate RCS measurement.}

\fs{Reverberation chambers (RCs) have been successfully used to achieve various types of electromagnetic characterization as detailed abundantly in the literature \cite{Corona,Book_hill,Book_RC,Book_RC_2}.}
An RC can be seen as an oversized electromagnetic cavity allowing the superposition of numerous modes with random complex weights according to the variation of boundary conditions. Mechanical, frequency or source stirrings (or a combination of them) yield to a statistically uniform and isotropic field. This test environment is of high interest for electromagnetic immunity tests since it provides high-field strength and does not require any movement of the equipment under test. Beyond electromagnetic compatibility applications, RCs are now used in many other domains as antenna characterization \cite{Antenna_measure,Antenna_measure_plane_wave,Antenna_SWD,Antenna_RC, Reis2022}, over-the-air testing of wireless devices \cite{OTA_1,OTA_2,OTA_3}, or exposure system for bio-electromagnetic studies \cite{Bio_RC,Bio_RC_mobile,Bio_RC_18}.

\fs{Although well suited to estimate average absorption quantities, some papers recently introduced novel approaches to estimate RCS patterns within RCs.}
To that purpose, different methods have been developed in order to extract the line-of-sight backscattered contribution of the target from the diffuse field provided by the RC. In \cite{K-factor}, the $K$-factor has been used in order to extract the line-of-sight signal between the transmitting and the receiving antennas from the scattered component associated to reflections and diffractions on RC walls. This direct path component has then been used to extract the RCS of a target \cite{RCS_K}. \cc{This method is relevant to  estimate high K-factor and therefore high RCS magnitude  but the RCS dynamic range is limited \cite{k_dir}}. Another approach consists in moving the target inside the RC in the direction of the measuring antenna, so that the backscatterred signal from the target can be associated to a virtually recreated Doppler effect \cite{Doppler}. \cc{It partially solves the problem of the $K$-factor method but requires a linear rail to alternately move the target closer and further away in the direction of the measuring antenna and for different viewing angles}. A time-gating method has been proposed in \cite{RCS_TG} which is similar to the current method used in an anechoic chamber and, therefore, not specific to RCs. In \cite{RCS_base}, some of the current co-authors have introduced a monostatic measurement principle based on a sinusoidal regression applied on the difference of the complex-valued frequency signals recorded in the empty chamber (calibration step) and the target-loaded chamber (measurement step). \fs{The theory has been further improved accounting for the mechanical stirrer effect and contribution in \cite{RCS_MS_RC} and extended to quasi-monostatic configurations in \cite{Eurad}}.

\fs{The current method is based on a sinusoidal regression process whose maximum amplitude versus the considered signal period leads to the target RCS. The regression is performed on a selected frequency bandwidth around the central frequency, and it has been demonstrated in \cite{RCS_MS_RC} that the larger the bandwidth, the higher the accuracy of the regression process itself. However, it is shown in this paper that a large frequency bandwidth degrades the accuracy of the retrieved RCS pattern by smoothing out the side lobes. Therefore, a two-step signal processing is introduced to identify the appropriate antenna-target distance on the one hand, and to improve the RCS amplitude estimation over all azimuth angles of the target, on the other hand.}

This manuscript is organised as follows. Section II briefly recalls the main hypotheses and the principle of RCS measurement in RCs as detailed in \cite{RCS_MS_RC}. Section III describes the  measurement environment, the target under study and also the data acquisition process. These measurement data are recorded in a quasi-monostatic configuration, at once with both antennas in horizontal (HH) and vertical (VV)  polarizations.  Then, section IV is dedicated to the impact of the frequency bandwidth selection on the estimation of the distance between the target and the antennas, whereas section V studies its impact on the RCS estimation itself. The two-step signal processing proposed in section VI is a direct consequence of the analysis in the two previous sections.  This advanced method is applied to the measured data in section VII highligthing a major improvement of RCS results if appropriate bandwidths are selected, specifically for local minima estimation of the  RCS pattern. Conclusions are drawn in Section VIII.

\section{Theory of RCS measurement in RCs}
In the following theoretical developments, the RC is supposed to be well oversized (with respect to the wavelength), so that Hill's hypothesis of an ideal random (diffuse) field can be assumed \cite{HILL PWS}. In practice, goodness-of-fit test provides a weak rejection of the null-hypothesis for a Gaussian field distribution. Furthermore, the target is assumed to behave as an equivalent scattering point-source.  The RCS pattern of such a target is evaluated from the difference of the scattering parameters measured by two fixed antennas in the RC without the target (thereafter called "empty room") and the RC loaded by the target for each azimuth position of the latter. The theory of RCS measurement within an RC is detailed in monostatic configuration in \cite{RCS_base,RCS_MS_RC} and has been extended to the quasi-monostatic configuration in \cite{Eurad}. The difference $\Delta S(f_0)$ of the transmission parameters measured within the charged room  $S_{xy}^\mathrm{T}(f_{0})$  and the empty room $S_{xy}(f_{0})$ writes for a given azimuth position of the target:
\begin{multline}
     \label{quasi-monostatic}
\Delta S(f_{0}) = S_{xy}^\mathrm{T}(f_{0})-S_{xy}(f_{0})=\\B(f_0)+ A(f_0)\times\exp\left(-\mathrm{j}\frac{2\pi f_{0}}{\delta f}\right)\times\exp(\mathrm{j}\phi_{0})
\end{multline}
with
\begin{multline}
     \label{B}
B(f_0)=\sqrt{m_x(f_{0}) m_y(f_{0})}\times\sqrt{\eta_{x}(f_0)\eta_{y}(f_0)}\\\times \left(H^\mathrm{T}(f_{0})-H(f_{0})\right)
\end{multline}
\begin{multline}
     \label{A}
A(f_0)=
\sqrt{\sigma^\mathrm{T}(f_{0})}\frac{c\sqrt{m_x(f_{0}) m_y(f_{0})G_{x}(f_{0})G_{y}(f_{0})}}{(4\pi)^{3/2}R^{\prime2}f_{0}}
\end{multline}
The VNA ports and related antennas are indicated by the $x$ and $y$ indexes, and the presence of the target by the superscript $\mathrm{T}$. The difference $\Delta S(f_0)$ consists of the addition of two distinct terms. The first term $B(f_0)$ on the right-hand side of (\ref{quasi-monostatic}) is related to the RC diffuse field. It is itself composed of three  factors as shown in (\ref{B}). The first two ones involve the free-space (FS) antenna mismatching coefficients $m_i(f_{0})=1-\lvert S_{ii\mathrm{FS}}(f_{0})\rvert^{2}$, and the antenna radiation efficiencies $\eta_{i}(f_0)$ with $i=\{x,y\}$, respectively. The last factor is proportional to the difference of the diffuse RC transfer functions with the target $(H^\mathrm{T}(f_{0}))$ and without the target $(H(f_{0}))$.  The second term $ A(f_0)\times\exp\left(-\mathrm{j}\frac{2\pi f_{0}}{\delta f}\right)\times\exp(\mathrm{j}\phi_{0})$ corresponds to the radar echo which is proportional to the square root of the RCS of the target $\sigma^\mathrm{T}(f_0)$, $A(f_0)$ being a real-valued parameter. The terms $G_{x}(f_{0})$ and $G_{y}(f_{0})$ in (\ref{A}) are the gains of the $x$ and $y$ antennas, respectively. The term $R'$ is the physical distance between the target and the antennas. In the frequency domain, this delayed echo can be approximated as a sinusoidal function around the central frequency $f_{0}$, whose periodicity $\delta f$  depends on the electrical distance $R$  with respect to the relation $\delta f=c/(2R)$. Finally, $\phi_{0}$ is considered as a constant phase. 

In our case, we further assume that the RC electromagnetic field statistical properties are not modified by the target. As a result, both variables  $H^\mathrm{T}(f_{0})$ and $H(f_{0})$ are distributed according to a centered Gaussian probability density function with the same variance. Therefore, the sinusoidal waveform is embedded in a Gaussian noise once the frequency variations of the antenna gains and  mismatch factors are compensated for as described in (\ref{SER_norm}).
\begin{equation}
\label{SER_norm}
\Delta S^\mathrm{N}(f_{0})=\Delta S(f_{0})\frac{(4\pi)^{3/2}R^{\prime 2 }f_{0}}
{c\sqrt{m_x(f_{0})m_y(f_{0})G_{x}(f_{0})G_{y}(f_{0})}}
\end{equation}
that can be decomposed as:
\begin{equation}
\label{SER_theo_decomposition}
\Delta S^\mathrm{N}(f_{0}) = D(f_0)+ \sqrt{\sigma^\mathrm{T}(f_{0})}\times\exp\left(-\mathrm{j}\frac{2\pi f_{0}}{\delta f}\right)\times\exp(\mathrm{j}\phi_{0})
\end{equation}

Through scattering parameters measurements over a certain bandwidth centered at $f_{0}$, applying a sinusoidal regression procedure on the difference $\Delta S^\mathrm{N}(f_{0})$ permits to extract the magnitude of the target RCS $\sigma^\mathrm{T}(f_0)$.
The magnitude of the sinusoidal wave is valued from the average of both amplitudes extracted from the real and the imaginary parts of $\Delta S^\mathrm{N}(f_{0})$. However, before estimating the sine component amplitude, we need to determine the distance $R$ that imposes its frequency period. As illustrated in Fig.~\ref{subfig_schematic}, this distance $R$ can be seen as the distance $R'$ between the target and the antenna apertures plus the electrical distance $\delta R$ between the antenna apertures and the setup calibration planes. In the earlier  publications \cite{RCS_base,RCS_MS_RC}, the amplitude $A(f_0)$ was estimated by fixing a distance $R$ whatever the target orientation. Nevertheless, in \cite{Eurad}, we highlighted that this distance varies according to the target orientation and we also introduced a method to estimate the distance between the target and the antennas by measuring the target response on a very large bandwidth.  Nevertheless, using a very large frequency bandwidth is not suitable for the RCS estimation as further highlighted in the following section. 

\section{Measurement Setup and Data Acquisition}
\subsection{Measurement Setup}
\begin{figure}[!t]
\subfloat[Picture of the measurement set up in HH polarization in the RC\label{subfig_picture}]{\includegraphics[width=\columnwidth]{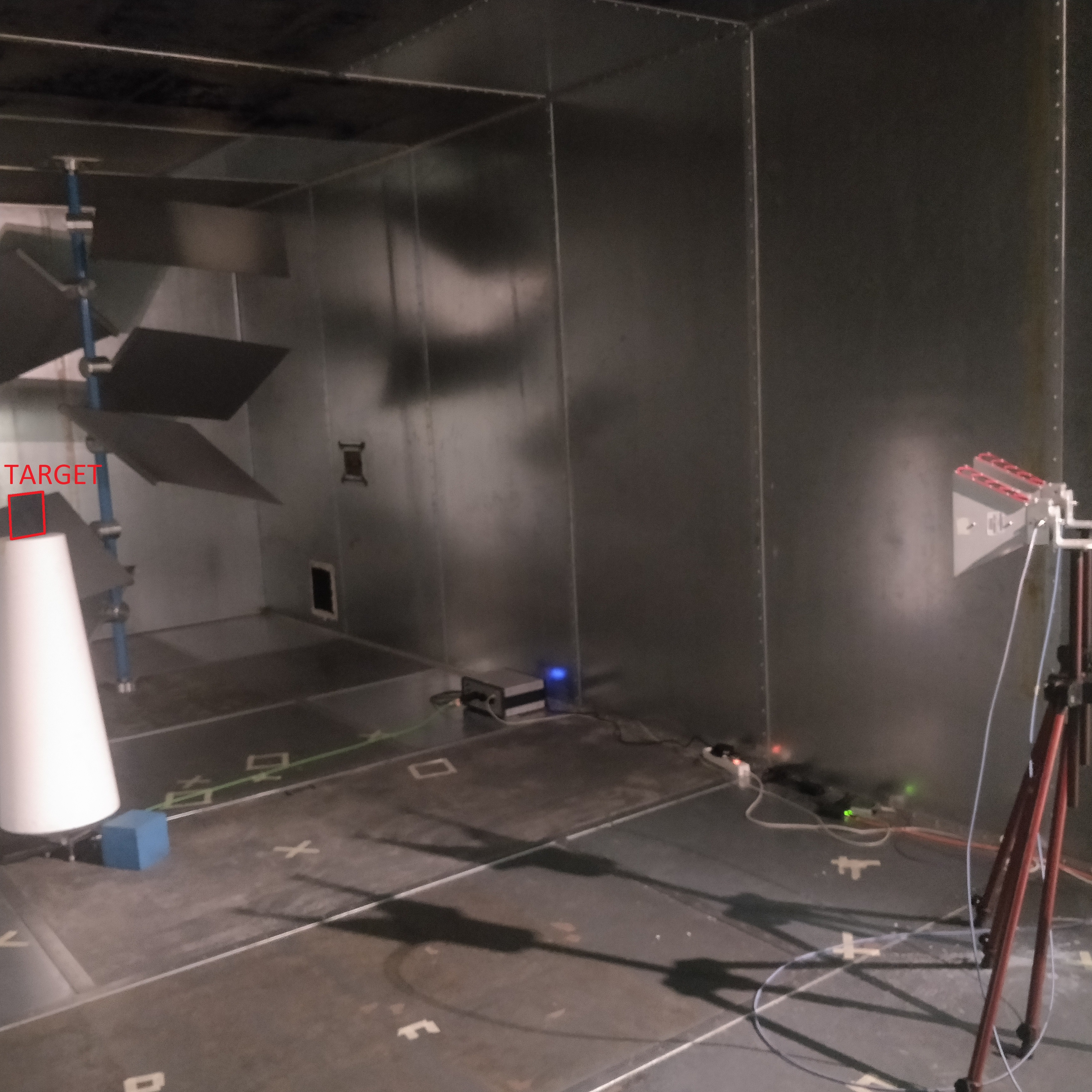}}
\\
\subfloat[Schematic description with distance definitions \label{subfig_schematic}]{\includegraphics[width=\columnwidth]{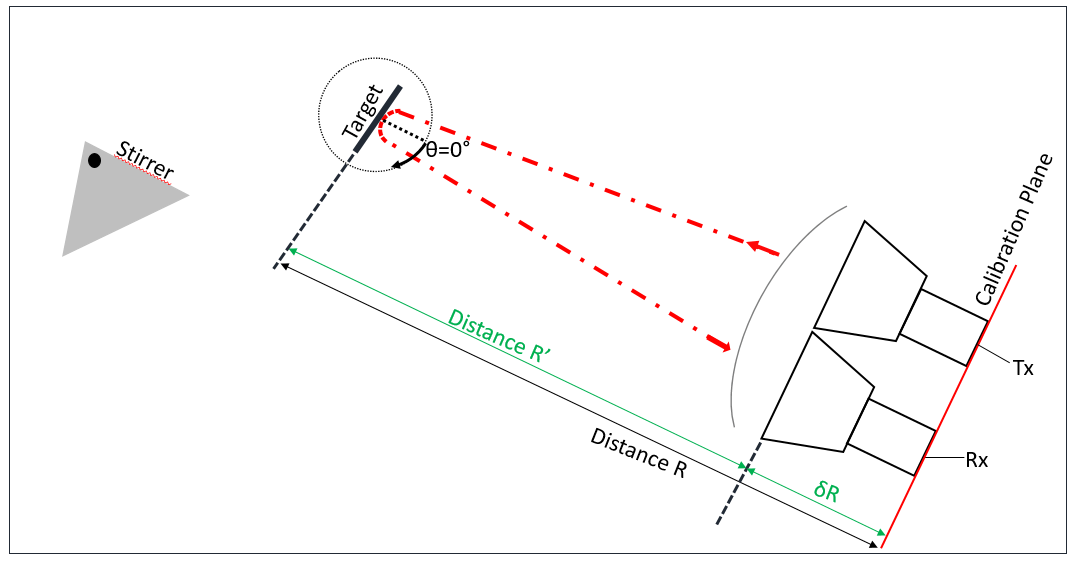}}
\caption{\cc{Picture in HH polarization (a) and schematic description with distance definitions (b) of the test setup with a metallic plate of $0.152\times 0.150$~m$^2$ in the RC.}}.
\label{photo_mesure}
\end{figure}

Measurements are made in the largest RC of the IETR. The dimensions of the chamber are $2.9$~m $\times$ $3.7$~m $\times$ $8.7$~m. The RC has a lowest usable frequency of $0.25$~GHz and behaves as an ideal RC above $0.8$~GHz if empty. It means that a goodness-of-fit test provides a weak rejection of the null-hypothesis for a Gaussian field above $0.8$~GHz. The RCS measurement is performed using two identical horn antennas (ETS-Lindgren model 3115) oriented towards the target and at a close distance from each other regarding the target point-of-view. Antennas are used in two different polarizations, the horizontal polarization (HH) and the vertical polarization (VV). \er{It has been verified that the coupling between the two measurement antennas due to the close distance between them does not affect the RCS measurement result \cite{Eurad}}. A rotating mast supporting the target is placed at the distance $R'=3$~m from both antennas (Fig.~\ref{subfig_picture}).
The orientation of the antenna-to-mast axis is chosen arbitrarily but not perpendicular to any cavity wall to avoid any spurious echo from the back wall behind the target. 

\subsection{Target under study and theoretical response}
The selected target is a rectangular metallic plate of length $l=0.152$~m and height $h=0.150$~m. Its theoretical RCS behavior is well-known (e.g. using physicalcal optic approximation \cite{RCS_OG,Ross} or geometrical diffraction theory \cite{Ross}). Thus, discarding the Gaussian noise of the RC, the theoretical response exhibits a wide RCS dynamic range over the azimuth plane. According to the physical optics approximation \cite{RCS_OG}, the monostatic target RCS $\sigma^\mathrm{T,Th}(f_0,\theta)$ in HH configuration as a function of the plate surface $lh$, the frequency $f_0$ and the azimuth position $\theta$, is given by: 
\begin{equation}
   \sigma^\mathrm{T,Th}(f_0,\theta)=\frac{4\pi(lh)^2 f_0^2}{c^2}\cos^2(\theta)\sinc^2\left(\frac{2\pi l \sin(\theta) f_0}{c}\right)
\end{equation}
As a consequence, the theoretical normalized $\Delta S^\mathrm{NTh}(f_0)$ expression for this plate discarding the Gaussian field in the RC ($B(f_0)=0$) writes:

\begin{equation}
\label{target}
\Delta S^\mathrm{NTh}(f_0) \approx \sqrt{\sigma^\mathrm{T,Th}(f_0,\theta)} \times  \exp{\left(\frac{-\mathrm{j} 2\pi\times f_{0}}{\delta f}\right)}
\end{equation}

\begin{figure}[!t]
\subfloat[$\theta=0^\circ$\label{subfig_0}]{\includegraphics[width=\columnwidth]{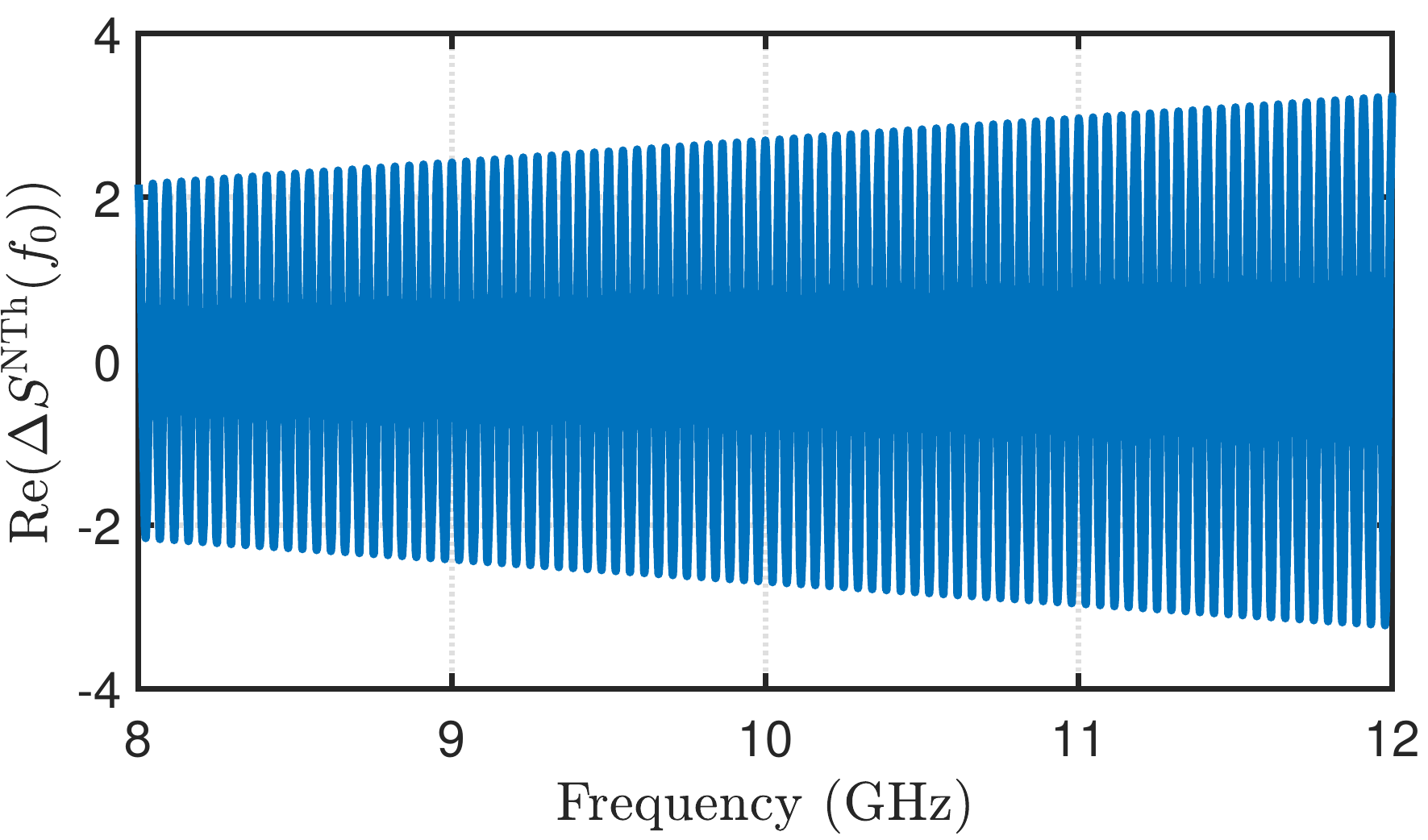}}
\\
\subfloat[$\theta=+35^\circ$\label{subfig_35}]{\includegraphics[width=\columnwidth]{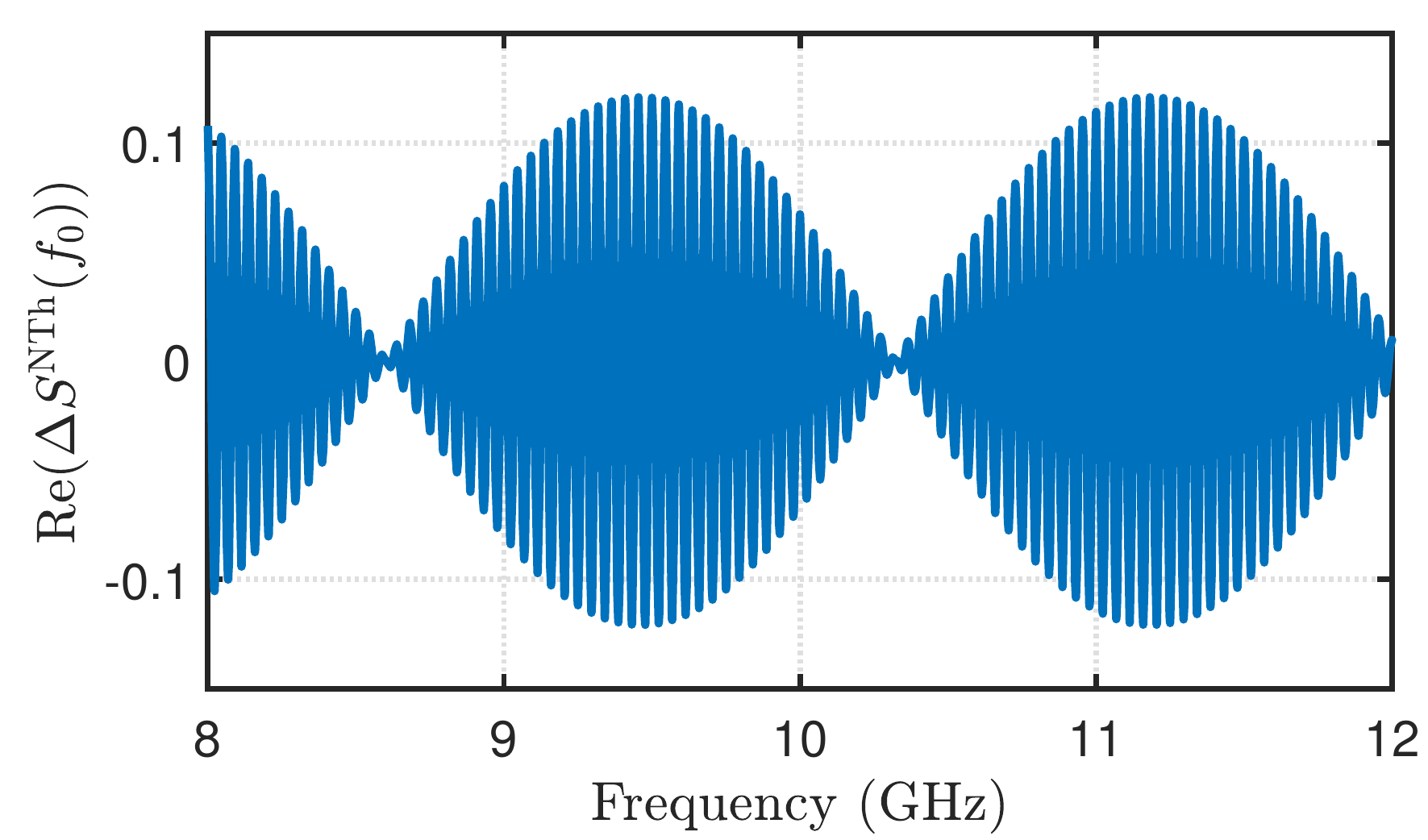}}
\caption{Real part of the theoretical noiseless difference of S-parameters ($\Delta S^\mathrm{NTh}(f_0)$) in absence and presence of a metallic rectangular plate at (a) $\theta=0^\circ$ and (b) $\theta=35^\circ$.}  
\label{Signal_theorique}
\end{figure}

The term $\Delta S^\mathrm{NTh}(f_0)$ phase shift variation versus frequency is written as a function of $\delta f = c/2R$ as described in (\ref{quasi-monostatic}). Fig.~\ref{Signal_theorique} shows the real part of (\ref{target}) over a bandwidth of $4$~GHz centered around $f_0=10$~GHz for $R=3.3$~m and for two angular positions of $\theta=0^\circ$ and $\theta=35^\circ$.

For $\theta=0^\circ$ (Fig.~\ref{subfig_0}), the amplitude of the sinusoidal wave raises up linearly with frequency. Due to this smooth variation, the  sinusoidal fitting in a least square sense would lead to similar amplitudes whatever the sub-bandwidth selected for this fitting.  On the contrary, in Fig.~\ref{subfig_35} ($\theta=35^\circ$), the amplitude of the noiseless  sinusoidal oscillations strongly varies over the frequency band and even vanishes at specific frequencies. Therefore, the amplitude extracted from the sinusoidal fit strongly depends on this frequency bandwidth. \sm{The latter must be restricted so that $\Delta S^\mathrm{NTh}(f_0)$ amplitude remains approximately constant to ensure a correct RCS estimation at angles for which the RCS variation is important versus frequency}. This paper aims at solving this issue of the suited frequency band selection over a large angle range as described in the next sections. 

\subsection{Data acquisition}
Both antennas are connected to a vector network analyzer (VNA). To respect the far field Fraunhofer distance, the distance $R'$ between the target under test and the antennas must be greater than $2 D^2/\lambda$  where $D$ is the largest dimension of the target and $\lambda$ is the minimum wavelength in the measurement bandwidth. As measurements are carried out from $8$~GHz to $12$~GHz, $R'$ is fixed at $3$~m and respects this limit. This condition is supposed to be sufficient to assume the ballistic wave being a locally-plane wave at the target position. A $1$~kHz IF filter and a frequency step of $100$~kHz are used ($40001$~frequency points in total). \fs{Indeed, the RC correlation bandwidth has been experimentally estimated to roughly $100$~kHz at $10$~GHz, hence the chosen frequency step.} The RC mechanical stirrer is maintained in the same position throughout all experiments (i.e. no mechanical stirring is used throughout this paper). Firstly, we carry out S-parameter measurement of the empty RC $(S_{xy}(f_{0}))$. Secondly, we add the target on the mast in the RC and we perform the measurement $(S^\mathrm{T}_{xy}(f_{0}))$ for  azimuth angles between $\theta = -100^\circ$ and $\theta = + 100^\circ$ with a $1^\circ$ rotation step.

\section{Distance estimation}
This section is dedicated to the accurate estimation of the distance between the target and the antennas. Using the algorithm described in \cite{Eurad}, the estimated distance is taken as the one leading to the maximum extracted sine wave amplitude over a distance range encompassing the total volume of the target. In our study, the distance (or the corresponding period of the sinusoidal fit) is varied  between $3.25$~m and $3.50$~m by a step of $0.001$~m. As the distance between both antennas is much smaller, the coupling between the two antennas has no effect on the distance estimation of the target. Thus, at each azimuth angle, the distance of the maximal amplitude is recorded as the optimal one. In the following, the effect of the frequency bandwidth used for the amplitude extraction is studied, comparing the optimal distance estimation for frequency windows of width varying from $\Delta f = 0.5$~GHz to $\Delta f = 4$~GHz, all centered at $f_0=10$~GHz. This study is performed for both HH and VV polarizations. 

\subsection{HH polarization}
\begin{figure}[!t]
\centerline{\includegraphics[width=\columnwidth]{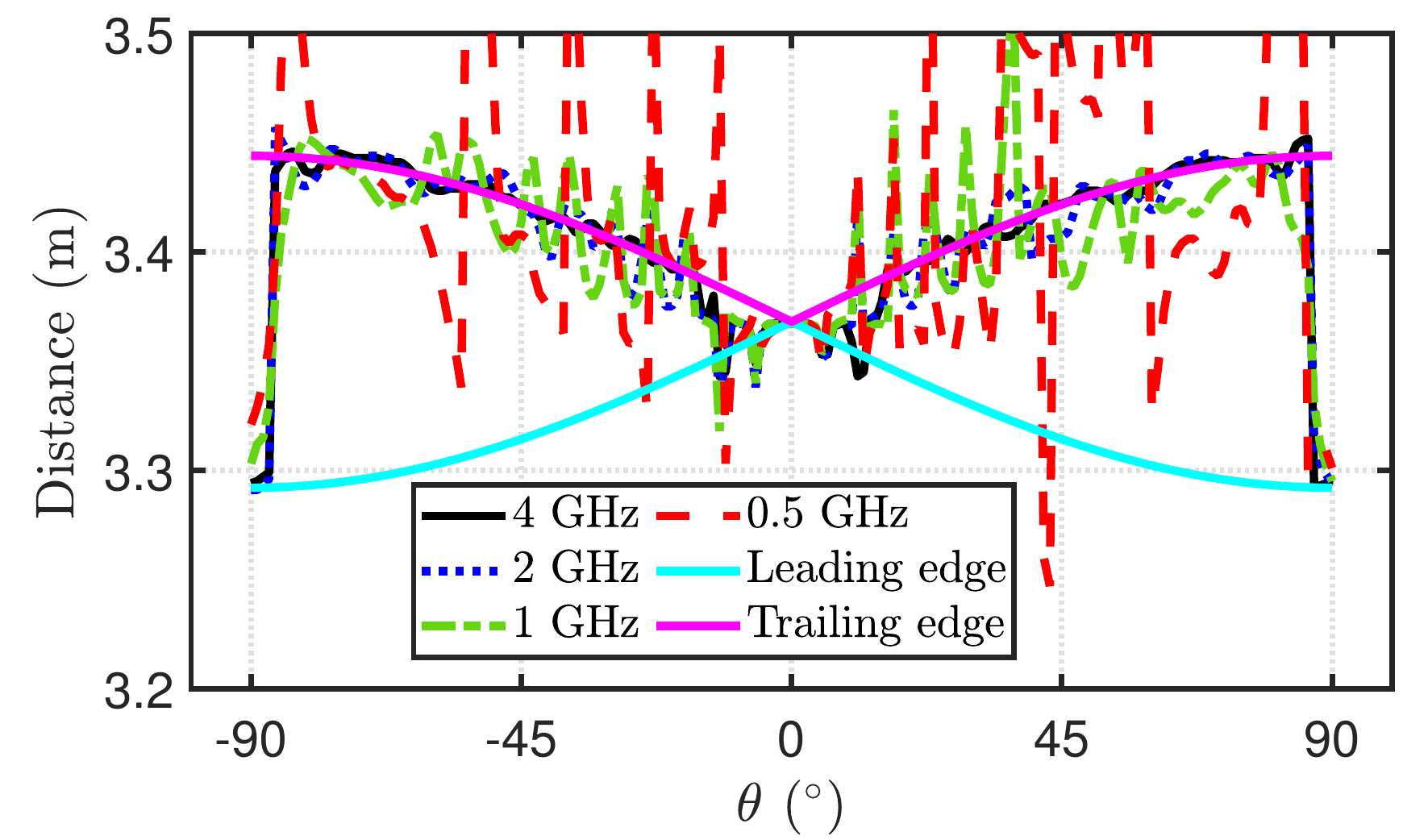}}
\caption{Distance evolution corresponding to the maximal amplitude of the sinusoidal regression (performed in the $3.25$~m to $3.50$~m range) as a function of the azimuth position $\theta$ and for different considered frequency windows in the HH configuration.}
\label{Distance_theta_HH}
\end{figure}

\er{Fig.~\ref{Distance_theta_HH} exhibits the estimated distance $R$ over the azimuth angles $\theta$ for different frequency bandwidths ($\Delta f=\{0.5,1,2,4\}$~GHz) in the HH configuration along with the theoretical positions of the plate leading and trailing edges}.
 Clear conclusions can be drawn. The distance estimated using the largest bandwidth ($4$~GHz) leads to the smoothest variation and the most physically coherent result. Indeed, the extracted distance is seen to coincide with either the trailing edge or, at some particular angles, the leading edge of the rotating metallic plate, that is consistent with the possible contribution of the varying current distribution on the plate. On the contrary, the distance is poorly estimated for the frequency window of $0.5$~GHz, exceeding the limits prescribed by the positions of the target leading and trailing edges. The two other frequency bandwidths ($1$~GHz and $2$~GHz) lead to intermediate estimations with less non-physical results than for $0.5$~GHz but not as accurate as for $4$~GHz. This is explained by the fact that the distance resolution $\delta d$ depends on the bandwidth $BW$ such as $\delta d = c/(2 \times BW) $. Therefore, the greater the bandwidth, the finer the resolution. Moreover, we can notice that the trailing edge is the main scattering part of the plate for $\lvert \theta \rvert \geq 10^\circ$, whereas the leading edge is dominant for $\lvert \theta \rvert < 10^\circ$. The main scattering point follows the theoretical position of the leading edge or the trailing edge. The abrupt change of distance between  $\theta=-90^\mathrm{\circ}$ and $\theta=-80^\circ$ is about $0.146$~m. As expected, it is very close to the length $l$ of the plate ($0.152$~m).

\begin{figure}[!t]
\centerline{\includegraphics[width=\columnwidth]{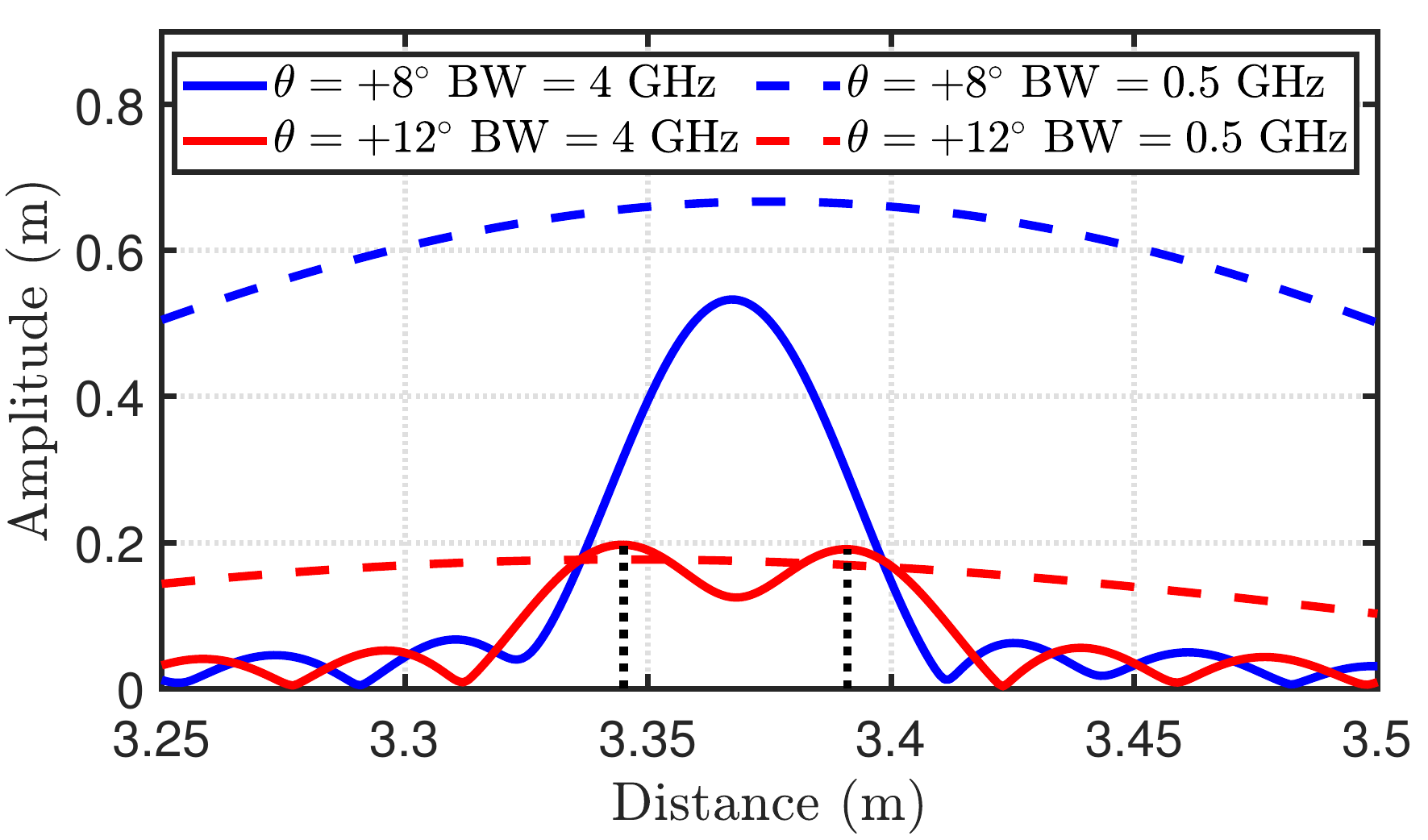}}
\caption{Amplitude estimated by the sinusoidal regression fitting in the $3.25$~to $3.5$~m distance range for $0.5$~GHz and $4$~GHz frequency windows and two different azimuth angles of the metallic plate in the HH measurement configuration. \pb{At $\theta=+12^\mathrm{\circ}$, the position of the two main scattering contributors, i.e. the leading edge and the trailing edge, are apparent.}}
\label{Amp_distance_Bande}
\end{figure}

These observations are further illustrated focusing on specific angles $\theta$ that correspond to local RCS minimum and maximum according to (\ref{target}). The amplitude of the sinusoidal regression fitting over the $3.25$~m to $3.50$~m distance range of HH polarization measurement results is plotted in Fig.~\ref{Amp_distance_Bande} for a local RCS maximum  ($\theta = +8^\circ$) and for a local RCS minimum ($\theta = +12^\circ$).

 Considering the local  maximum RCS located at $\theta = +8^\circ$, Fig.~\ref{Amp_distance_Bande} shows that the distance is much more precisely assessed since the frequency bandwidth is large. Then we are able to precisely localize the target area which mainly contributes to the backscattered echo. Around this local RCS maximum contribution, the distance is correctly identified with the largest bandwidth. On the contrary, the conclusion is different for the local minimum RCS at $\theta = +12^\circ$. As shown in Fig.~\ref{Amp_distance_Bande}, two main scattering contributions of the plate are obviously visible located at $\theta  = +12^\circ$ for the frequency bandwidth of $4$~GHz and that is not the case for the frequency bandwidth of $0.5$~GHz. For this latter, the estimated distance is the mean distance between the two edges and the antennas. For $\theta  = +12^\circ$, the target may be represented by the contribution of two scattering zones. \cc{The two maxima obtained with a bandwidth of $4$~GHz are located at $3.345$~m and  $3.391$~m and are therefore $0.046$~m apart. The theoretical distance of the leading edge at the same angle is  $3.352 $~m \cite{Eurad}. This is close to the estimated distance. As far as the trailing edge is concerned, its theoretical distance is $3.383$~m. Once again, it corresponds approximately to the second maximum of the curve}. If we now consider paths associated to both estimated distances (corresponding to the path from the antenna to one target edge then back to the receiving antenna), the difference between the phase shifts of the traveling waves is of $\Delta \phi= \frac{2\pi \times 0.046}{\lambda} \approx 3.1 \pi$. As a consequence, both contributions are approximately out of phase, as expected for a RCS minimum.  

\begin{figure}[!t]
\centerline{\includegraphics[width=\columnwidth]{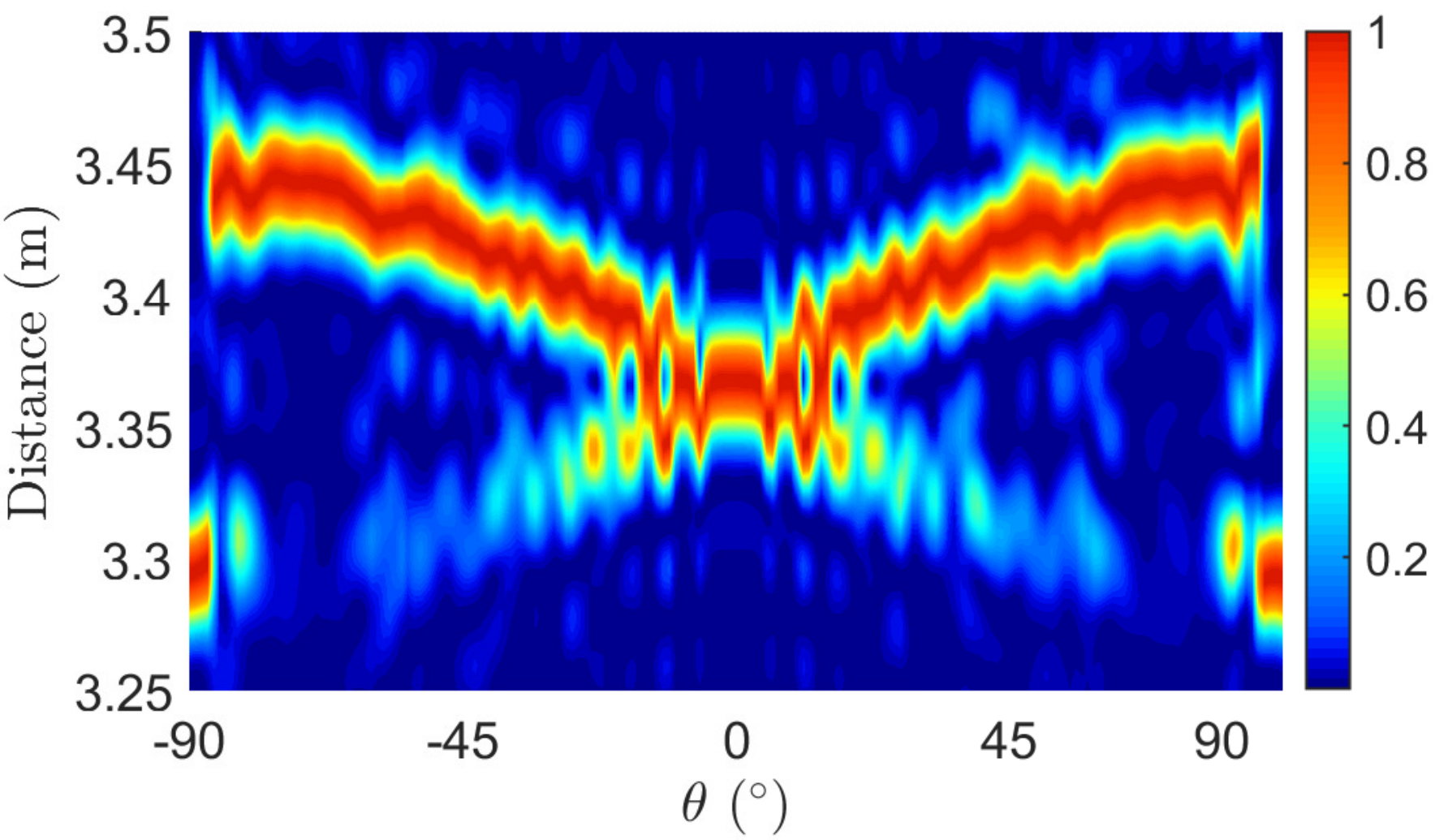}}
    \caption{Amplitude  of the sinusoidal regression fit in the $3.25$~to $3.50$~m range as a function of the azimuth angle ($-90^\circ \leq \theta \leq +90^\circ$) in the HH polarization considering the largest frequency bandwidth ($4$~GHz). The amplitude is normalized to its maximal value at each angle.}
\label{3D_HH}
\end{figure}

Fig.~\ref{3D_HH} presents the fluctuating amplitude provided by the sinusoidal regression fit at each distance and each azimuth position $\theta$ with the largest frequency bandwidth ($4$~GHz). 
For each azimuth position $\theta$, that corresponds to the variation of the amplitude estimated by the sinusoidal regression fit at a fixed azimuth position for a distance varying from $3.25$ to $3.50$~m as in Fig.~\ref{Amp_distance_Bande}. The amplitude is normalized by its maximum value at each angle enabling a clear visualization of the distance evolution over all azimuth positions of the target. The cartography exhibited in Fig.~\ref{3D_HH} highlights the location of the highest amplitude (equal to 1) of the sinusoidal regression for each value of $\theta$. The trailing edge position has a significant contribution for all angles  whereas the leading edge contributes only at specific angles. Regarding the position of the regression maximal amplitude, a  similar pattern to that of Fig.~\ref{Distance_theta_HH} is obtained.

\subsection{VV polarization}
\begin{figure}[!t]
\centerline{\includegraphics[width=\columnwidth]{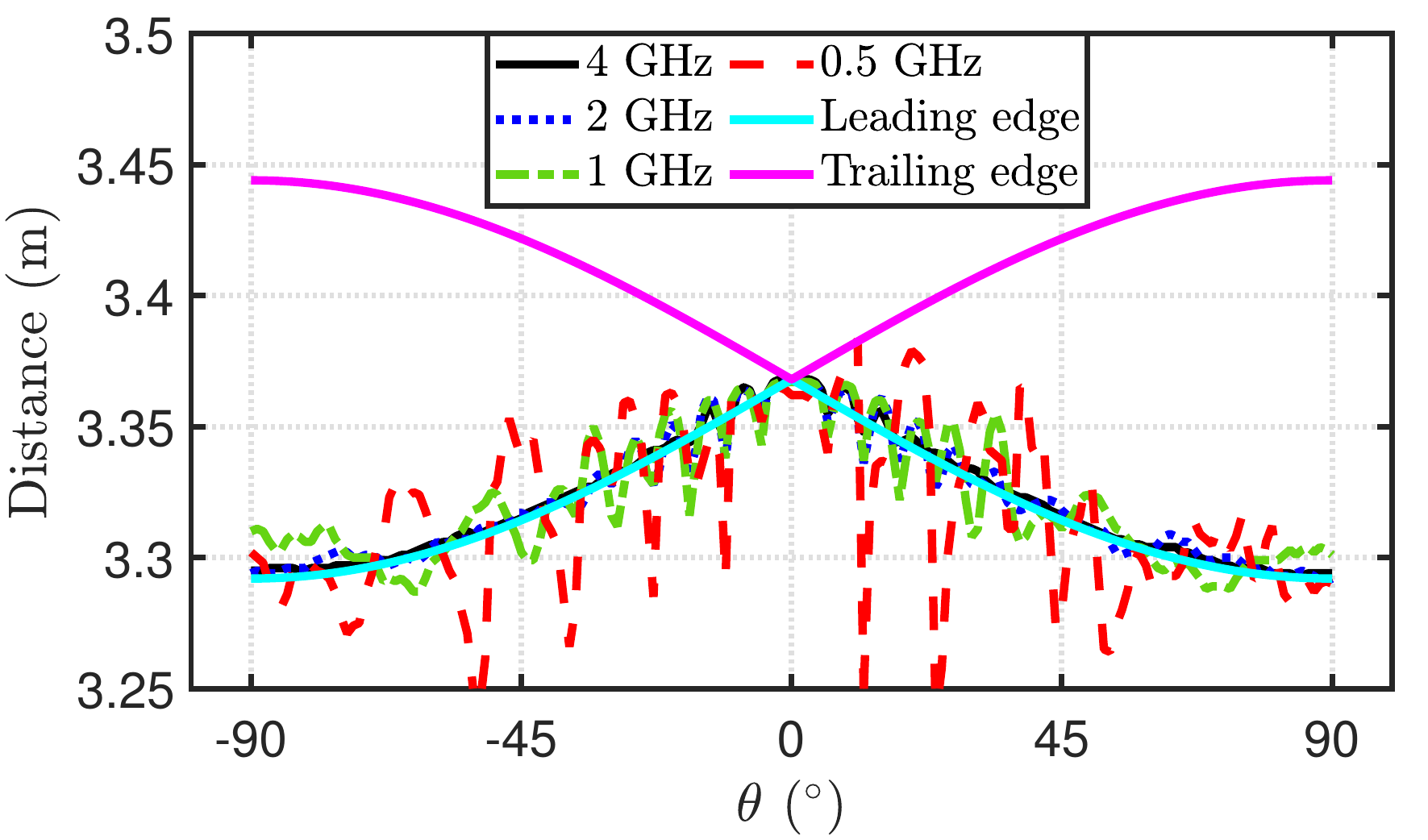}}
\caption{Distance evolution corresponding to the maximal amplitude of the sinusoidal regression (performed in the $3.25$~m to $3.50$~m range) as a function of the azimuth position $\theta$ and for different considered frequency windows in the VV configuration.}
\label{Distance_theta_VV}
\end{figure}

\begin{figure}[!t]
\centerline{\includegraphics[width=\columnwidth]{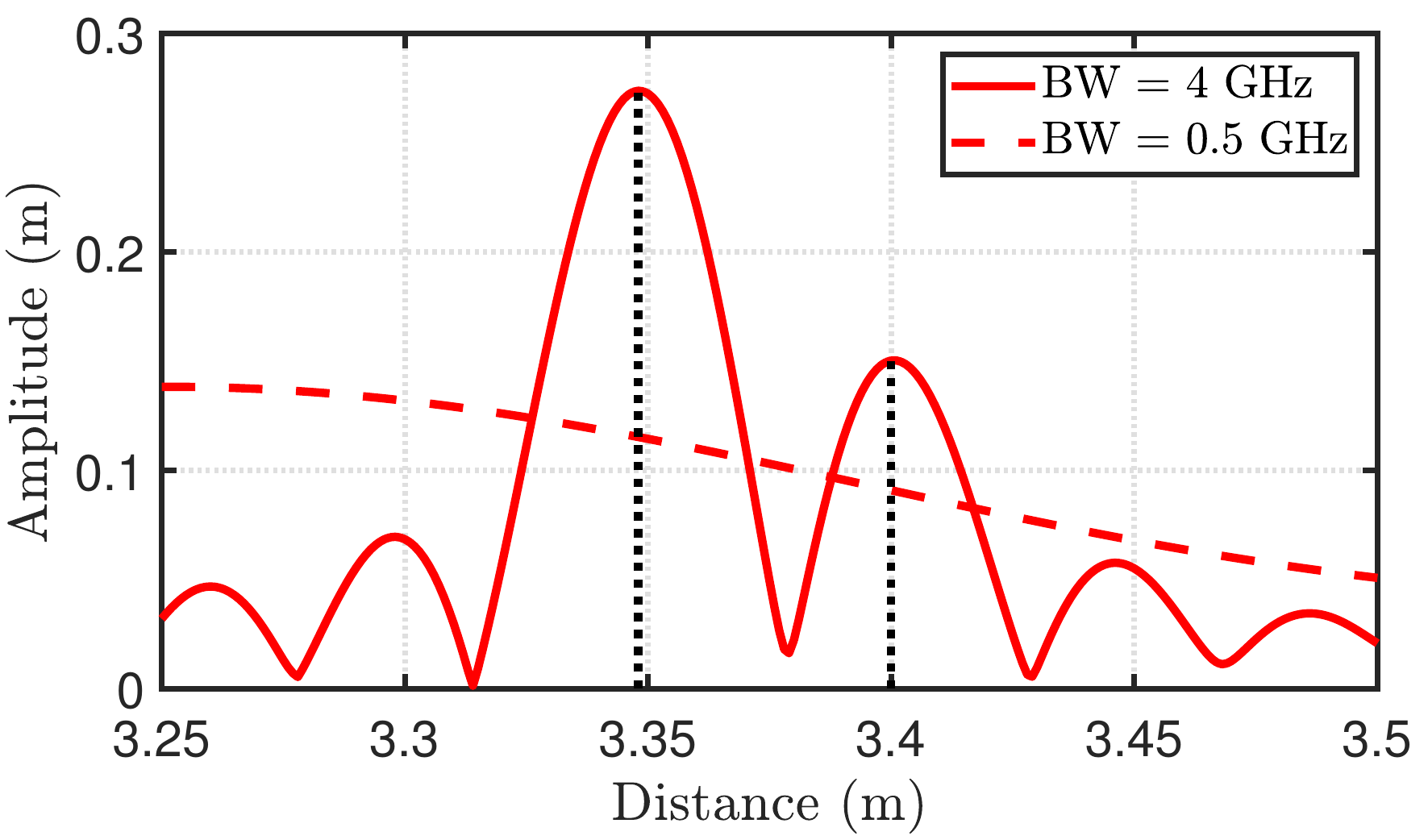}}
\caption{Amplitude estimated by the sinusoidal regression fit in the $3.25$~m to $3.5$~m distance range for $0.5$~GHz and $4$~GHz frequency windows in the VV configuration when the plate azimuth angle is of $12^\circ$. At that position, we can see the position of the main scattering contributions i.e. the predominant leading edge on the left and the less contributing trailing edge on the right.}
\label{amp_dist_VV}
\end{figure}

\begin{figure}[!t]
\centerline{\includegraphics[width=\columnwidth]{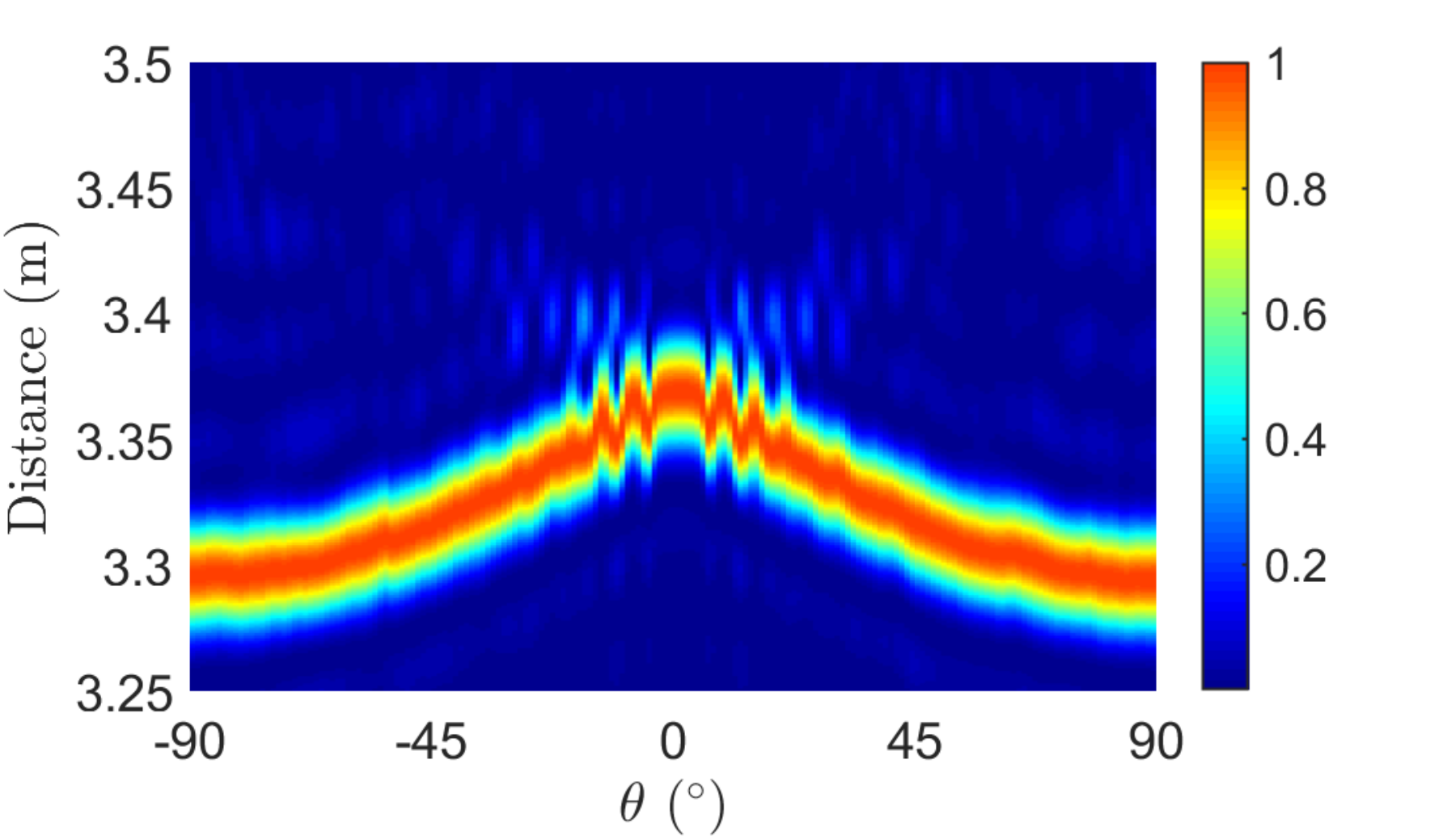}}
\caption{Amplitude  of the sinusoidal regression fit in the $3.25-3.50$~m range as a function of the azimuth angle ($-90^\circ \leq \theta \leq +90^\circ$) in the VV polarization considering the largest frequency bandwidth ($4$~GHz). The amplitude is normalized to its maximal value at each angle.}
\label{3D_VV}
\end{figure}

The same analysis is carried out in the VV configuration. The distance evolution of the maximal amplitude issued from the sinusoidal fit is shown for different frequency bandwidths ($\Delta f=\{0.5,1,2,4\}$~GHz) in Fig.~\ref{Distance_theta_VV}.
On the one hand, the VV configuration differs from the HH one since the leading edge constitutes the major contributor to the backscattering wave. On the other hand, like in the HH measurement configuration, only the large frequency bandwidths make possible to retrieve the theoretical distance of the leading edge. As in the HH polarization, the  difference between the distances estimated at $\theta=0^\circ$ and $\theta=+90^\circ$ is about $7.5$~cm, that corresponds to the half-plate length. Besides, it remains possible to identify the position of the two edges from the amplitude evolution of the sinusoidal fit as a function of distance. This is clearly visible in  Fig.~\ref{amp_dist_VV} for the azimuth position of $\theta = + 12^\circ$.
The two edges are indeed identified like in the HH configuration by the two amplitude maxima. However, the two edges are not associated to the same amplitude in the VV polarization, the leading edge is predominant. Fig.~\ref{3D_VV} is equivalent to Fig.~\ref{3D_HH} in the VV configuration and provides a normalized cartography of the fluctuating amplitude  of the sinusoidal regression fit in the same conditions. The leading edge is predominant and the trailing one is barely visible.

\subsection{Summary and milestone conclusion}
The target RCS amplitude is provided through a sinusoidal regression fitting applied on (\ref{SER_theo_decomposition}). As the regression fitting applies for a sine wave whose period depends on the distance between the antennas and the major backscattering contributor, the accurate estimation of this distance $R$ is a prerequisite for the signal amplitude extraction. We have shown here that it is not possible to estimate correctly the distance with a small frequency bandwidth and that the best distance estimation is obtained with a frequency bandwidth of $4$~GHz whatever the polarization configuration. Therefore, from now on, the largest frequency bandwidth of $4$~GHz is systematically used to estimate the distance between the target and the antennas. 

\begin{figure}[!t]
\centerline{\includegraphics[width=\columnwidth]{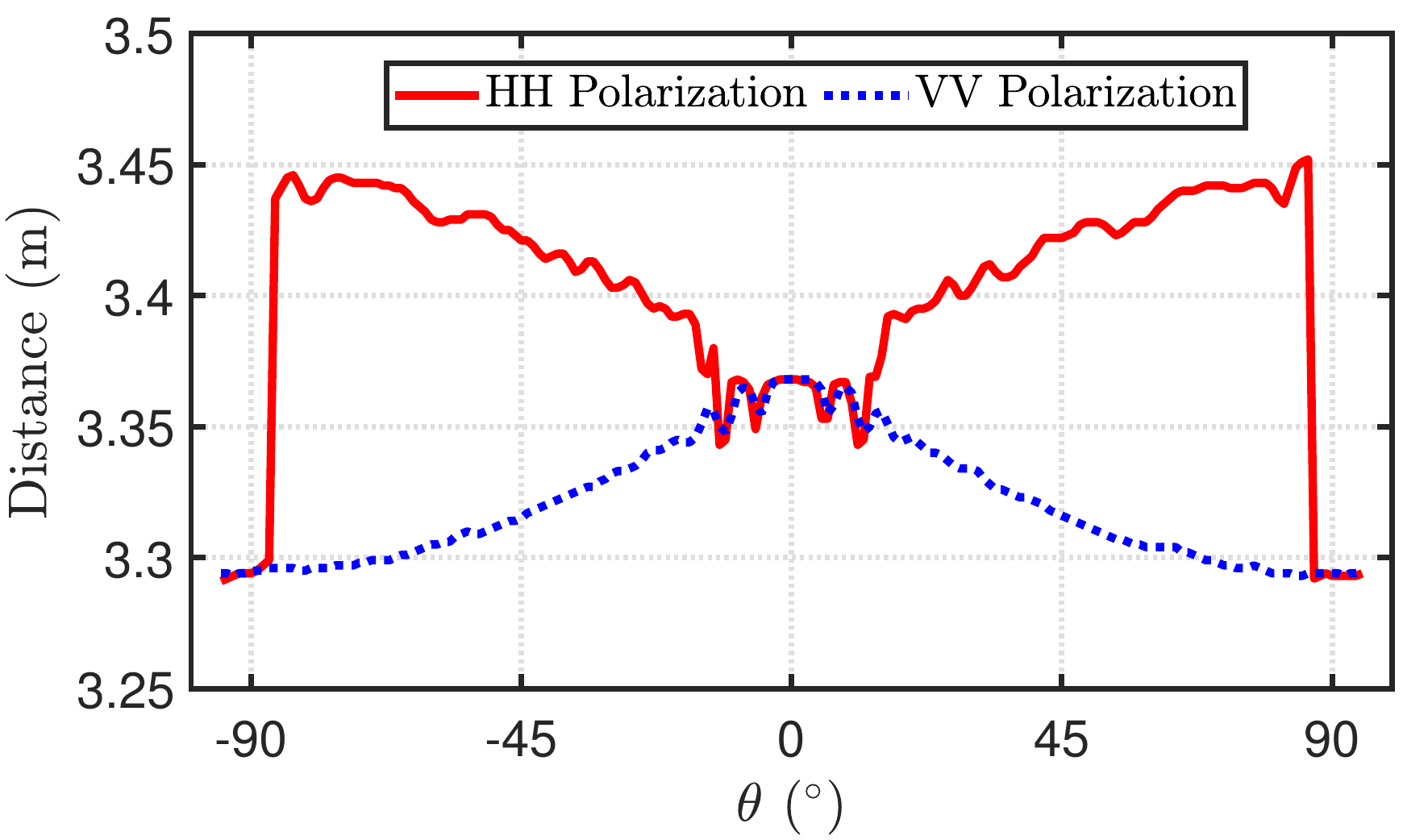}}
\caption{Comparison of the estimated distances corresponding to the maximal amplitude of the sinusoidal regression fit on a bandwidth of $4$~GHz as a function of the target azimuth angle $\theta$ varying from $-90^\circ$ to $+90^\circ$ for the two different measurement configurations.}
\label{Distance_theta_HHvsVV}
\end{figure}  

In Fig.~\ref{Distance_theta_HHvsVV}, we exhibit the distance estimated in HH and VV configurations using the $4$~GHz bandwidth and illustrate the position of the main scattering point in each polarization for different target azimuth angles. The distances retrieved in both configurations are similar for $\theta$ around $0^\circ$ and $\pm 90^\circ$. \sm{Apart from these angular ranges, the major contributor is the leading edge for the HH polarisation and is the trailing edge for the VV polarisation}.

\section{Impact of the frequency bandwidth on RCS estimation}
\begin{figure}[!t]
\centerline{\includegraphics[width=\columnwidth]{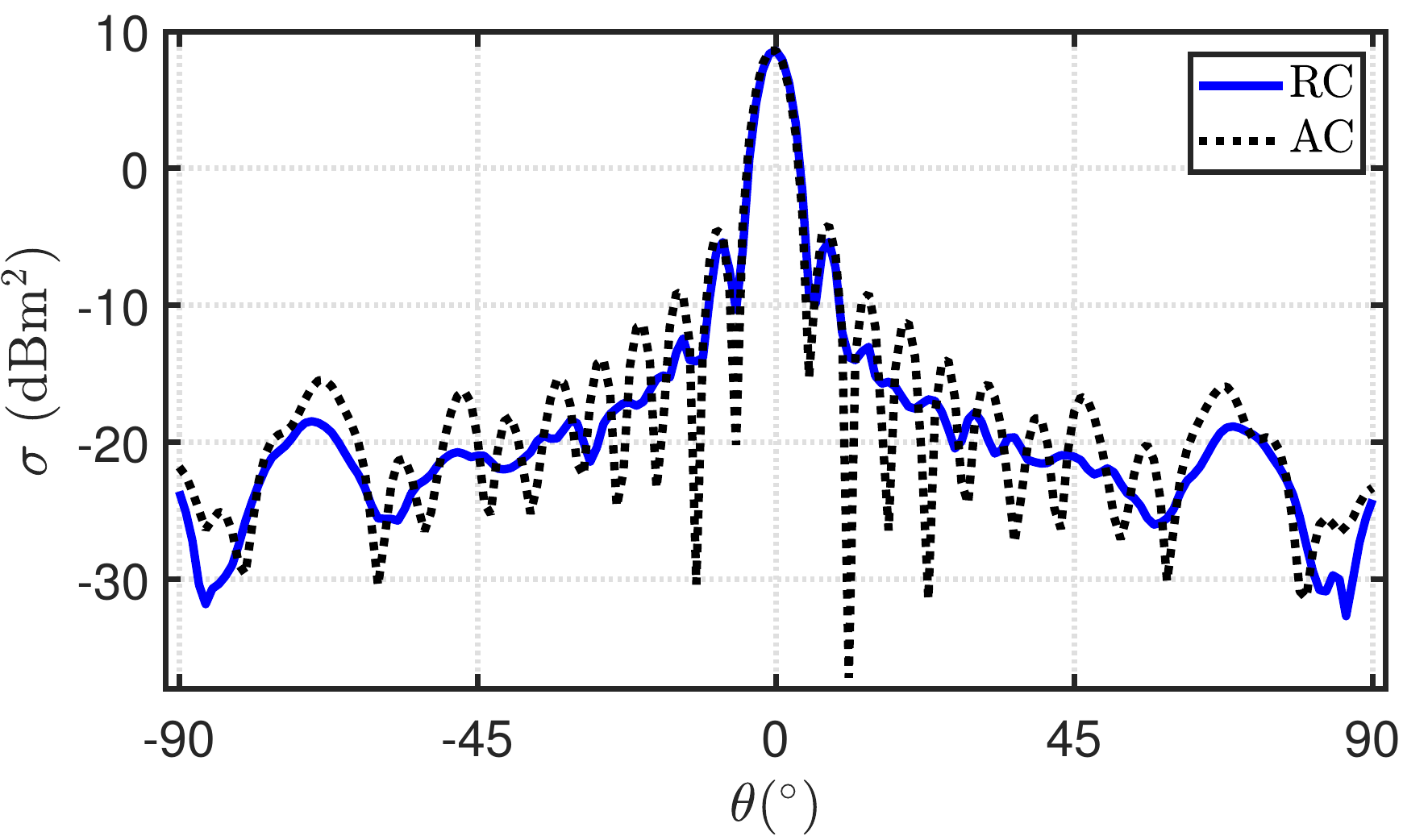}}
\caption{\cc{RCS evolution as a function of $\theta$ in the range from $-90^\circ$ to $+90^\circ$: comparison between the RCS extracted from RC measurement using a $4$~GHz window for the distance estimation as well as the amplitude fitting and the RCS measured in an AC.}}
\label{Comp_RCS_RC}
\end{figure}

\begin{figure}[!t]
\centerline{\includegraphics[width=\columnwidth]{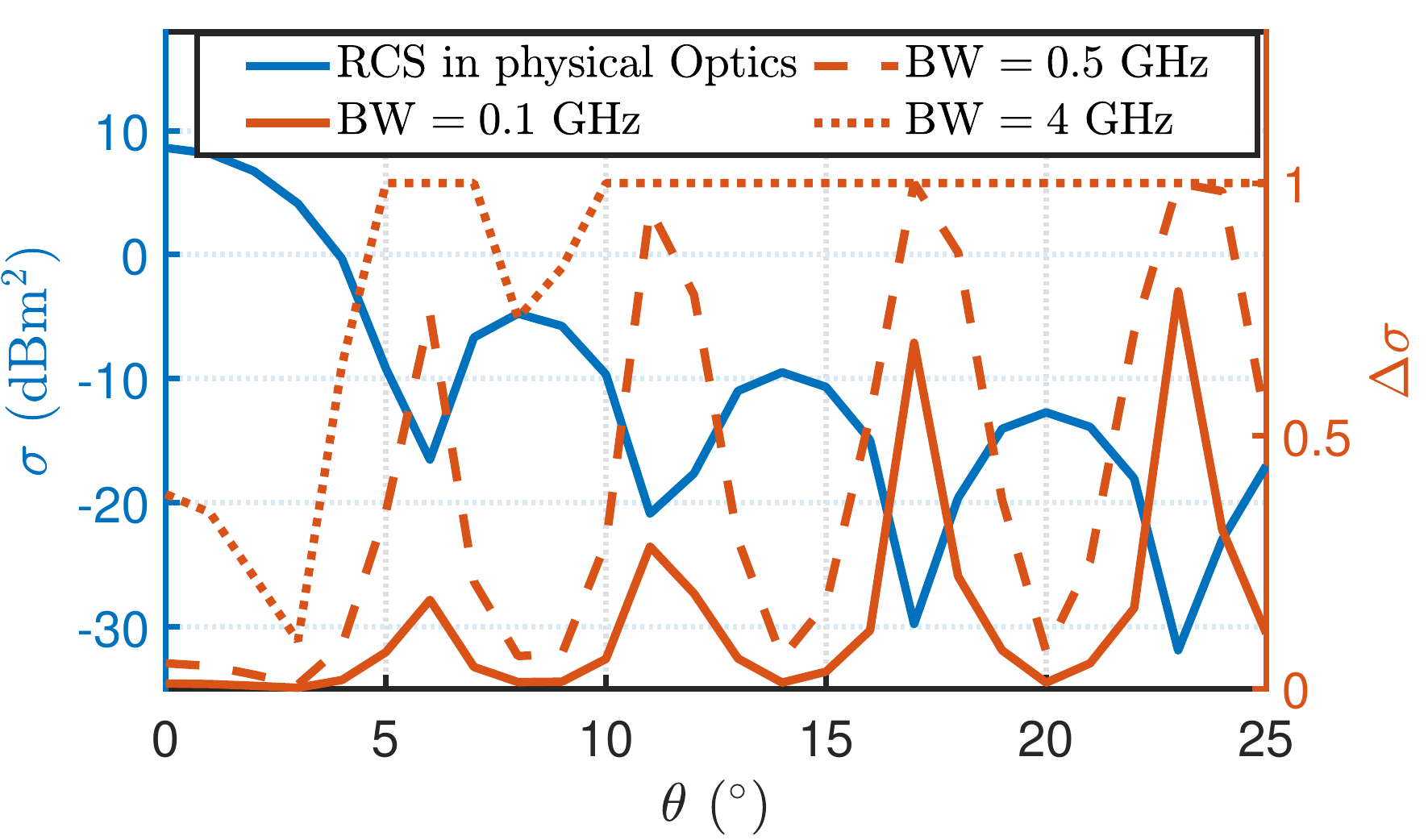}}
\caption{Theoretical RCS (PO formulation) and the relative error evolution w.r.t. the azimuth orientation $\theta$ in the range from $0^\circ$ to $25^\circ$ and according to different frequency bandwidths.}
\label{difference_f}
\end{figure}

In the previous section, the selection of the largest frequency bandwidth has been shown to be the better choice to  accurately estimate the antenna-target distance, i.e. the periodicity of the sinusoidal waveform to be fitted. This section is dedicated to the frequency bandwidth analysis for estimating the RCS amplitude itself, once the scattering point distance is known. First, we extract the RCS by using the same frequency bandwidth of $4$~GHz. In Fig.~\ref{Comp_RCS_RC}, we exhibit a comparison between the RCS extracted from measurements in  RC using a $4$~GHz bandwidth for distance and amplitude extractions and a measurement of the RCS in an AC.
Although there is a good agreement between both estimations around the RCS main lobe, the RCS extracted from RC measurements is smoothed and fails to discriminate the local minima and maxima. Consequently, the use of a large bandwidth is not efficient to estimate the RCS pattern. Based on Fig.~\ref{Signal_theorique}, this result could be anticipated as the RCS amplitude $\sigma(f_0,\theta)$ fluctuates versus frequency in a non-monotonic way. Given $\theta$ and $f_0$, the lower the frequency bandwidth, the smaller the fluctuations of the sine wave amplitude. Consequently, a better accuracy of the RCS amplitude is expected. To quantify the frequency variation of the target RCS according to the chosen frequency bandwidth ($BW$), we use the analytical physical optics (PO) expression of the plate RCS and compute the RCS excursion over $BW=\{0.1,0.5,4\}$~GHz at $f_{0}= 10$~GHz as follows:  

\begin{equation}
\centering
\label{delta_sigma_formule}
\Delta \sigma(f_0,\theta) =\frac{\underset{df}{\text{max}}(\sigma (f,\theta))-\underset{df}{\text{min}}(\sigma (f,\theta))}{\underset{df}{\text{max}}(\sigma (f,\theta))+\underset{df}{\text{min}}(\sigma (f,\theta))}
\end{equation}
with $df=[(f_0-\frac{BW}{2});(f_0+\frac{BW}{2})]$.
\newline
\newline
In Fig.~\ref{difference_f}, the plate RCS is plotted in blue at the central frequency $f_0 = 10$~GHz and for $0^\circ \leq \theta \leq 25^\circ$.
This figure highlights the increase of RCS fluctuations according to (\ref{delta_sigma_formule}) for increasing bandwidths. This RCS excursion is bounded between $0$ and $1$. The relative min/max difference is especially important at angles of RCS local minima where RCS frequency derivative is high. Thus, the RCS is better estimated for a small frequency bandwidth. However, as described in \cite{RCS_MS_RC}, the number of uncorrelated realizations must be high enough for a correct estimation of the RCS amplitude. A trade-off must be found to ensure both low RCS fluctuation and low amplitude estimation uncertainty. Therefore, the two-step method presented in the next section is introduced to solve this issue. 

\section{The two-step method}
\begin{figure}[!t]
\centerline{\includegraphics[width=\columnwidth]{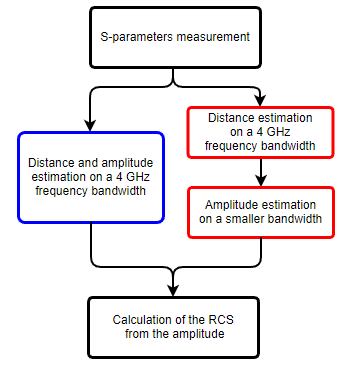}}
\caption{\pb{Flow-chart of the one-step method (left) and of the two-step method (right) in RC.}}
\label{Algo}
\end{figure}

In the previous sections, we have shown that the most suited frequency bandwidth choice is different for the estimation of the distance between the target and the antennas and the extraction of the backscattering signal amplitude leading to the RCS value. A two-step approach is thus proposed to account for these different requirements. It consists of: 
\begin{enumerate}
    \item Estimating the distance (i.e., the periodicity of the sinusoidal waveform to be fitted). This step is performed using data over the whole measurement bandwidth, i.e. $4$~GHz in our case as determined in section IV. It is called the distance bandwidth in the rest of this paper.
    \item Extracting the sine-wave amplitude through a regression fit, while restricting the processed data to a smaller bandwidth, called the amplitude bandwidth in the rest of this paper.  
\end{enumerate}
\cc{The two different post-processing methods used in RCs are summarized in the Fig~\ref{Algo}.} 

The reduction of the amplitude bandwidth aims at limiting the variation of the target RCS highlighted in Section V. Moreover, a sufficient number of data is necessary in the fitting step for an accurate amplitude extraction. To determine the amplitude bandwidth leading to the most accurate RCS extraction, the antenna-target distance is firstly estimated using a $4$~GHz bandwidth; then, the RCS of the target is extracted for various frequency bandwidths ranging from $0.05$~GHz to $4$~GHz in order to find the best trade-off. \er{Its impact on the amplitude estimation is observed through the relative difference between the RCS values obtained from RC and AC measurements}.  
 
\subsection{HH Polarization}
\begin{figure}[!t]
\centerline{\includegraphics[width=\columnwidth]{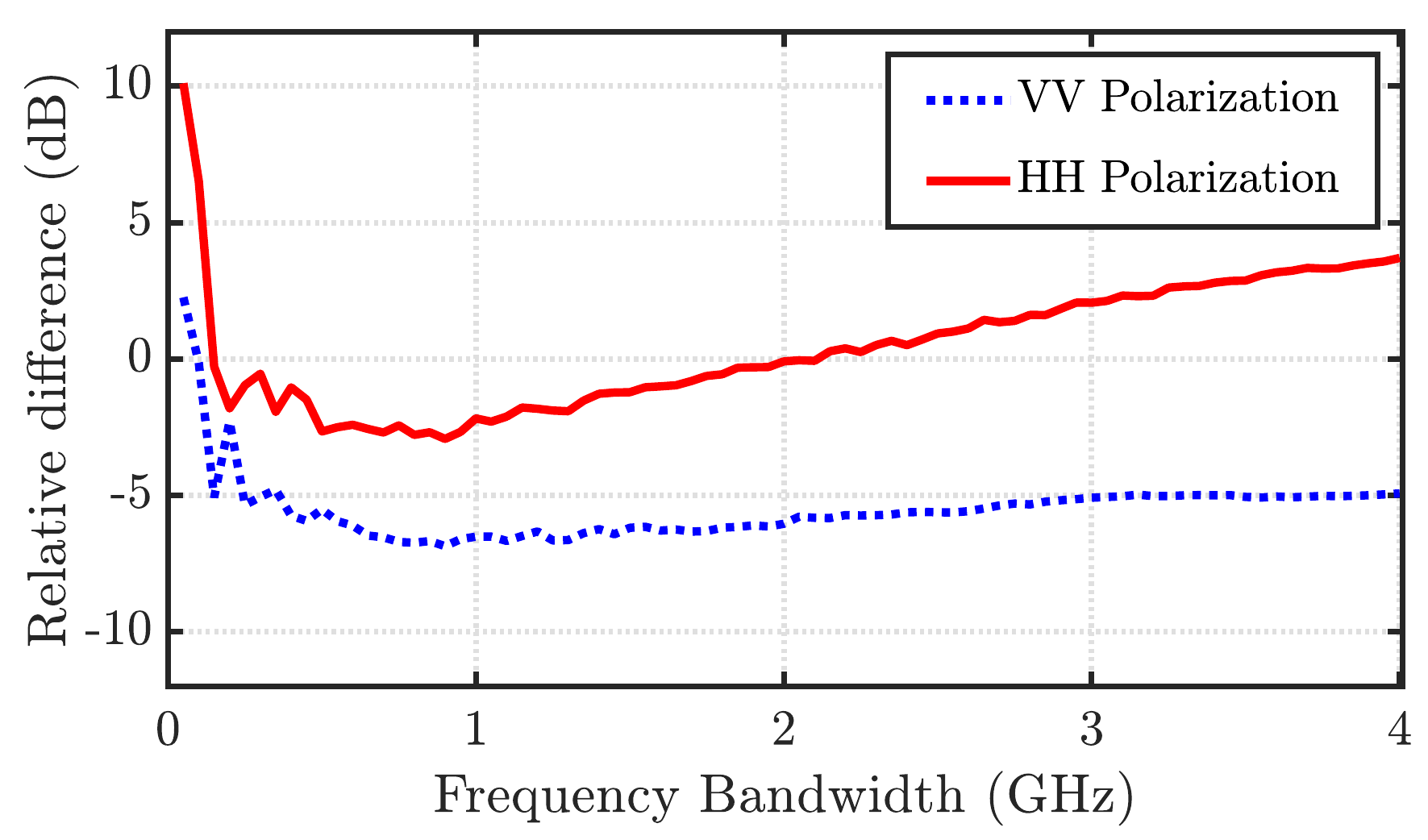}}
\caption{\er{Relative difference between extracted RCS values from RC and AC measurements in the HH and VV polarizations versus the frequency bandwidth used for the sine-wave amplitude extraction. Amplitude bandwidths ranging from 0.05 to 4~GHz are considered.}}
\label{error}
\end{figure}

First RC measurements are performed with both horn antennas in HH polarization and the extracted RCS values are studied for various amplitude bandwidths. \cc{ Fig.~\ref{error} shows the evolution of the difference between RCS values obtained from RC and AC measurements according to the amplitude bandwidth chosen to extract the sine wave amplitude. This average relative difference $\Delta$ is computed as follows:}

\begin{equation}
\centering
\label{Relative Difference}
\Delta=\left<\frac{\left|\sigma_\mathrm{AC}-\sigma_\mathrm{RC}\right|}{\sigma_\mathrm{AC}}\right>
\end{equation}
\\

Fig.~\ref{error} highlights that the relative difference evolution is not monotonic as it tends to decrease for amplitude bandwidths from 0.05~GHz to about 0.9~GHz and then to slightly increase for amplitude bandwidths from 0.9~GHz up to 4~GHz. \er{The decrease at the lowest bandwidth range is explained by a too low number of considered noisy data during the regression process. As already shown in \cite{RCS_MS_RC}, a sufficient number of uncorrelated realizations is necessary to extract accurately the amplitude, leading to a large enough frequency bandwidth. After this decrease, the slope of the curve changes and the relative difference increases with the frequency bandwidth. This behaviour is linked to the frequency variation of the RCS  and thus of  the sine wave amplitude.} From Fig.~\ref{error}, the best choice for the  amplitude bandwidth is situated in the interval from $0.5$~GHz to $1.5$~GHz. \cc{It has to be underlined that this indicated interval is specific to the RC in which the measurements are carried out and to the RCS of the target used in this measurement. However, a rectangular metallic plate exhibits large frequency fluctuations and can be considered as a limit case usable for the definition of the two-step method and particularly of the upper frequency value of this interval. The lowest frequency value strongly depends on the number of independent realizations of that RC or, in other words, on the RC correlation bandwidth}. Here the $0.5$~GHz bandwidth corresponds approximately to a set of $5000$ independent realizations.   

\subsection{VV Polarization}
In this section, the same study is performed with both horn antennas in the VV polarization. The evolution of the relative error according to the amplitude bandwidth is presented in Fig.~\ref{error} with the blue dotted curve. \er{As in the HH polarization case, the fitting process requires a sufficient number of uncorrelated realizations to extract accurately the RCS;  therefore the relative difference between AC and RC results is the largest with the smallest frequency bandwidth}. Then, the relative error decreases as the frequency bandwidth increases reaching a minimum around $0.9$~GHz. Beyond, the relative error goes up. The result is less impressive but consistent with that obtained in the HH polarization. This difference is only due to the smallest fluctuation of the RCS over the azimuth angle in the VV polarization. \er{As a consequence, the smoothing effect on the extracted amplitude of a large frequency bandwidth is less harmful to the RCS accuracy}. The conclusion drawn for HH configuration still holds: the amplitude bandwidth should be located in the range from $0.5$~GHz to $1.5$~GHz.

\section{Final RCS Results}
In this section, the two-step method is applied using a $4$~GHz distance bandwidth and $0.9$~GHz amplitude bandwidth. \cc{This smaller bandwidth is an appropriate choice to minimize the relative difference between AC and RC results with our RC and the considered rectangular metallic plate target}.

\subsection{HH Polarization}
\begin{figure}[!t]
\centerline{\includegraphics[width=\columnwidth]{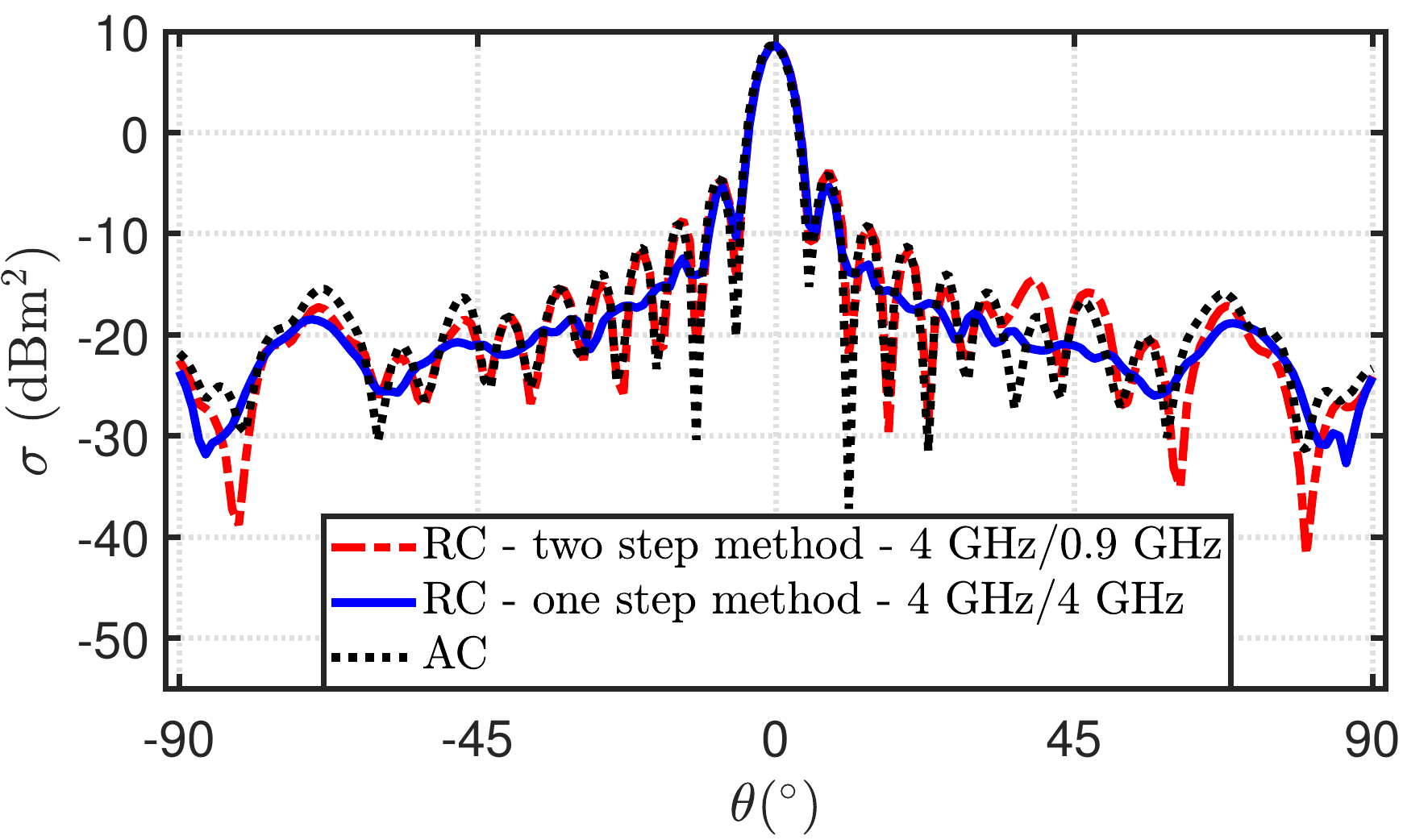}}
\caption{\pb{RCS extracted using a distance computed on a $4$~GHz window and a sine-wave amplitude estimated using a $0.9$~GHz bandwidth (two-step method) or a $4$~GHz bandwidth (one-step method) versus the plate  position in the $-90^\circ$ to $+90^\circ$ azimuth range and in the HH polarization.}}
\label{RCS_Bande}
\end{figure}

The two-step method is compared to the single-step one, i.e. choosing the same bandwidth of $4$~GHz for both distance and amplitude estimations. Fig.~\ref{RCS_Bande} presents the RCS in the HH polarization obtained with two different bandwidths along with the one measured in the AC (considered here as a reference).
\fs{As expected, the result obtained with the one-step method
does not replicate the fluctuations of the reference RCS (AC measurement), i.e., the RCS side lobes are smoothed out.
On the contrary, the RCS obtained with the two-step method
better agrees with the reference RCS, enabling to retrieve local maxima and minima. 
Therefore, it is demonstrated here that the two-step method increases the accuracy of the RCS estimation within RCs. To quantitatively compare the two RCS estimations within the RC, we compute the average relative difference $\Delta$ according to (\ref{Relative Difference}). An average relative difference of $234\% $ is reached between AC and RC measurements with a frequency bandwidth of $4$~GHz (one-step method). This average relative difference is strongly reduced down to $50\%$ with the smaller frequency bandwidth (two-step method). This remaining error can be mainly explained by larger relative errors at low RCS values and around local minima.} 

\subsection{VV Polarization}

\begin{figure}[!t]
\centerline{\includegraphics[width=\columnwidth]{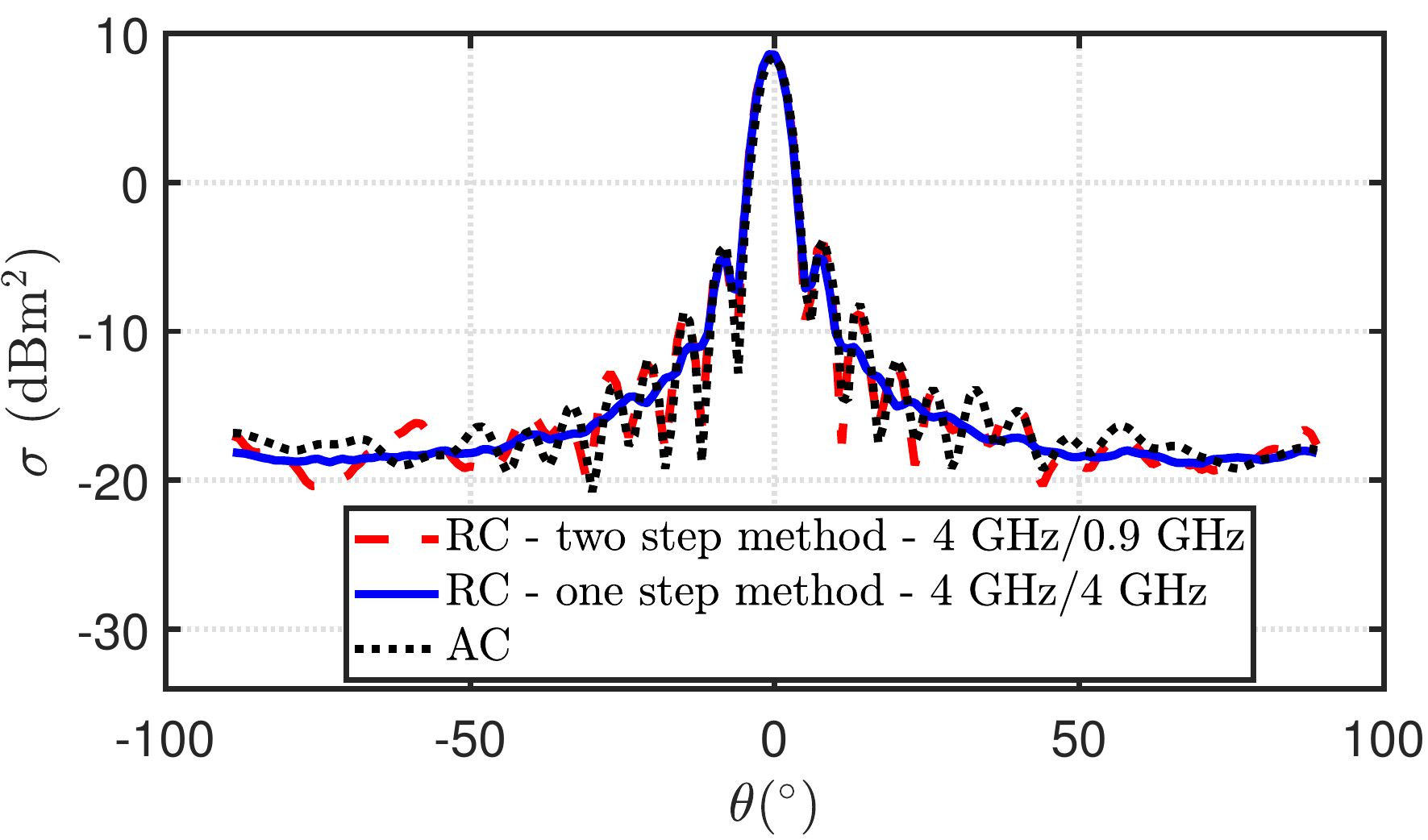}}
\caption{\pb{RCS extracted using a distance computed on a $4$~GHz window and a sine-wave amplitude estimated using a $0.9$~GHz bandwidth (two-step method) or a $4$~GHz bandwidth (one-step method) versus the plate  position in the $-90^\circ$ to $+90^\circ$ azimuth range and in the VV polarization.}}
\label{RCS_Bande_VV}
\end{figure}

The two-step method is now applied to the VV configuration in the same way. \cc{Once again, Fig.~\ref{RCS_Bande_VV} highlights the improvement of the extraction accuracy with the two-step method. Whereas the relative difference is of $32\%$ with the one-step method, it is decreased to $20\%$ with the two-step method. As already observed, the differences are smaller in this polarization as the minimal RCS values are larger than in the HH polarization. Thus, Fig.~\ref{RCS_Bande_VV} shows a satisfying agreement with the RCS measured in AC if the  RCS amplitude is computed on a small frequency bandwidth of $0.9$~GHz.}
 Besides, it has to be noticed that the relative difference is increased by the small angular offset observed on the measurements in regard to RCS maximum location.

\subsection{RCS extraction at different frequencies $f_0$}

\begin{figure}[!t]
\subfloat[RCS at $9$~GHz in HH polarization\label{RCS_9GHz_HH}]{\includegraphics[width=\columnwidth]{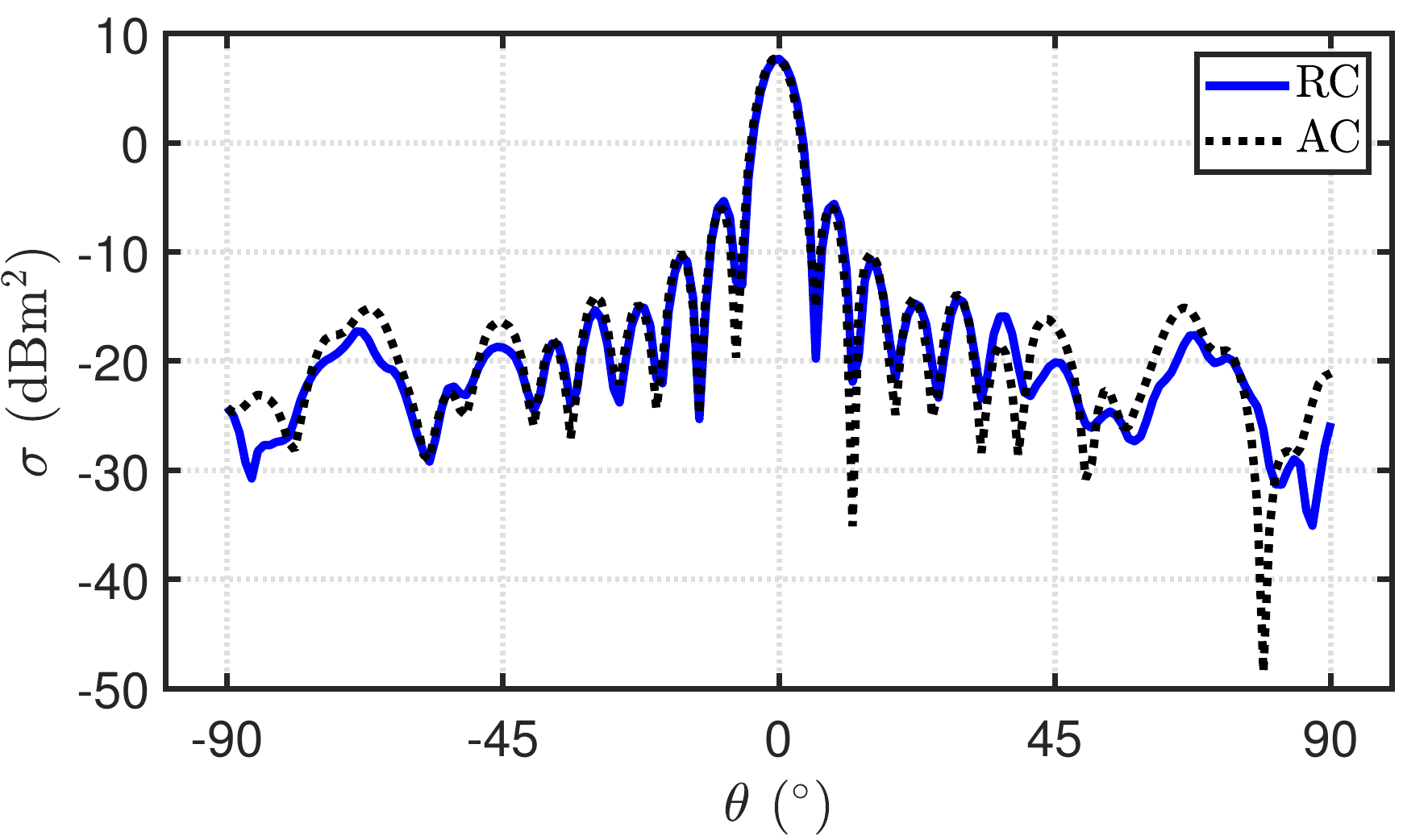}}\\
\subfloat[RCS at $11$~GHz in HH polarization\label{RCS_11GHz_HH}]{\includegraphics[width=\columnwidth]{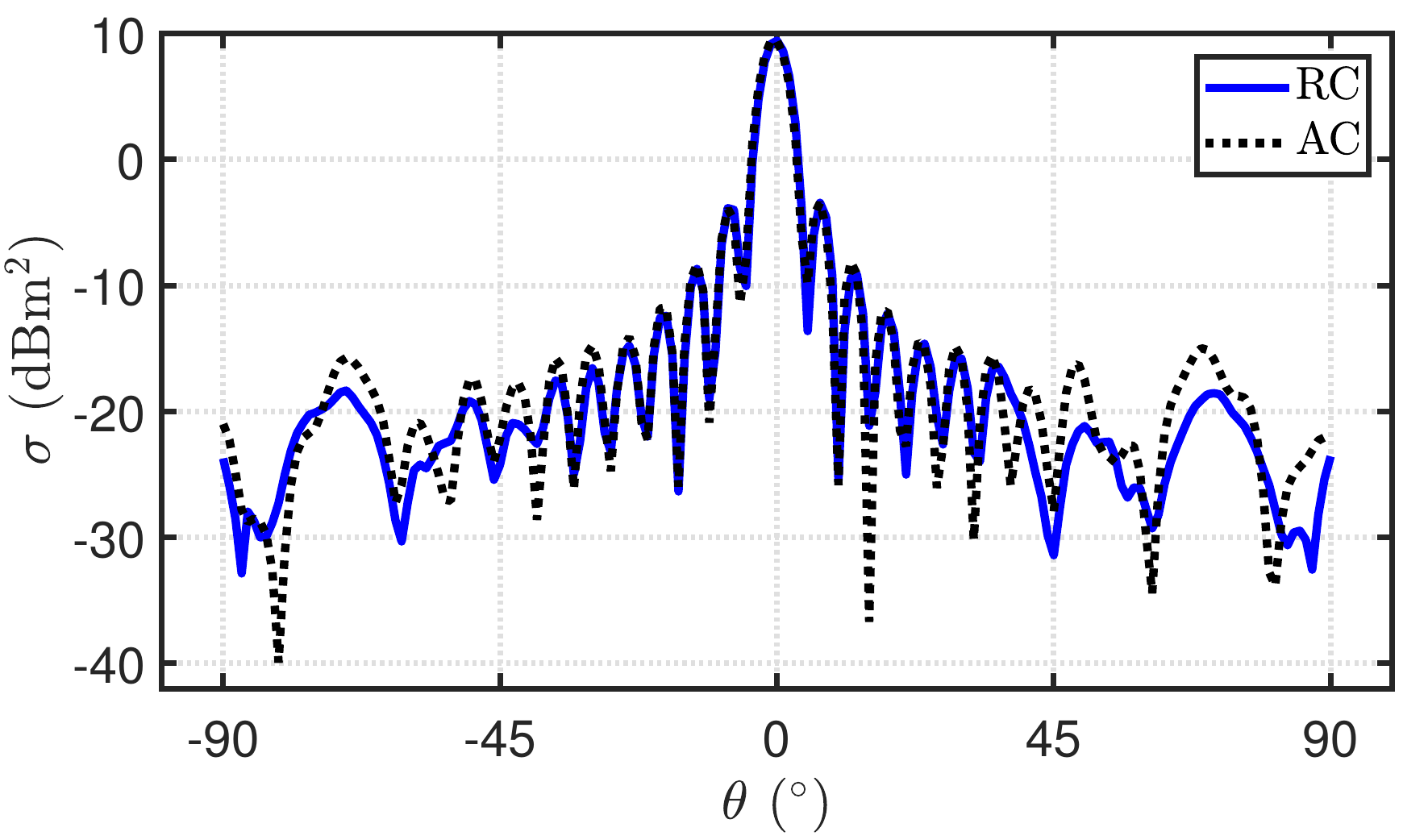}}\caption{\pb{Comparison of RCS extracted from RC measurements using a distance bandwidth of $4$~GHz between $8$~GHz and $12$~GHz and an  amplitude bandwidth of $0.9$~GHz for different central frequencies ($9$~GHz and $11$~GHz) and RCS obtained from AC measurements according to the plate  position in the $-90^\circ$ to $+90^\circ$ azimuth range and in the HH polarization.}}
\label{difference_f_HH}
\end{figure}

The developed method is now used to extract from the same measurement data the RCS at different frequencies, that are  $9$~GHz and $11$~GHz. First, we extract the distance (or period) using the large  bandwidth of $4$~GHz between $8$~GHz and $12$~GHz. Then, the amplitude of the sinusoidal regression fit is estimated with a frequency bandwidth of $0.9$~GHz centered around $9$~GHz on the one hand or around $11$~GHz on the other hand. The HH polarization results are shown in Fig. \ref{difference_f_HH} and are also compared with the AC measurement.
These results indicate that the RCS estimation can be performed in RC  for arbitrary values of the central frequency, using a limited frequency windows for amplitude estimation. The obtained results show that it is technically feasible to estimate the RCS pattern over the $9$~GHz and $11$~GHz frequency ranges.

\begin{figure}[!t]
\subfloat[RCS at $9$~GHz in VV polarization\label{RCS_9GHz_VV}]{\includegraphics[width=\columnwidth]{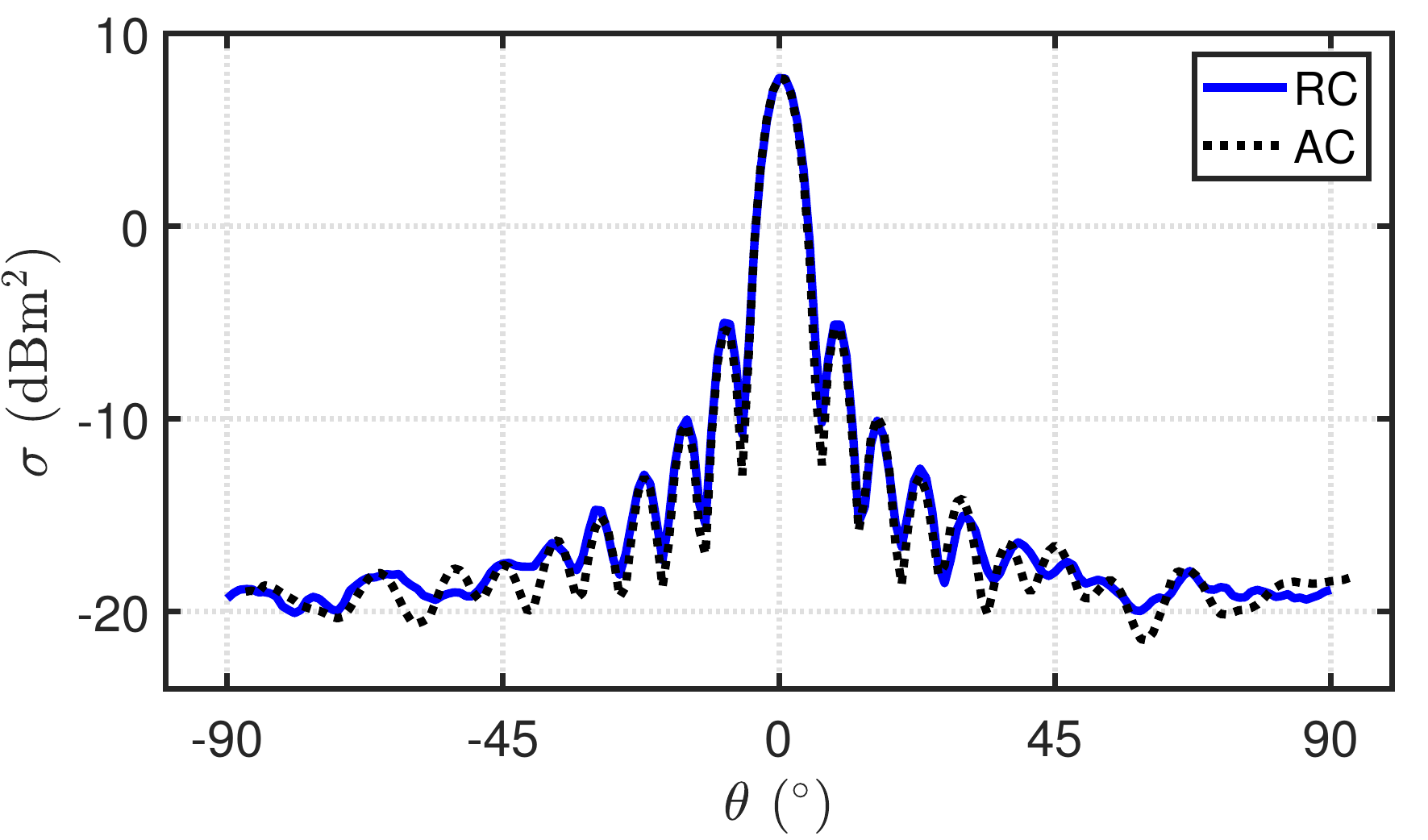}}\\
\subfloat[RCS at $11$~GHz in VV polarization\label{RCS_11GHz_VV}]{\includegraphics[width=\columnwidth]{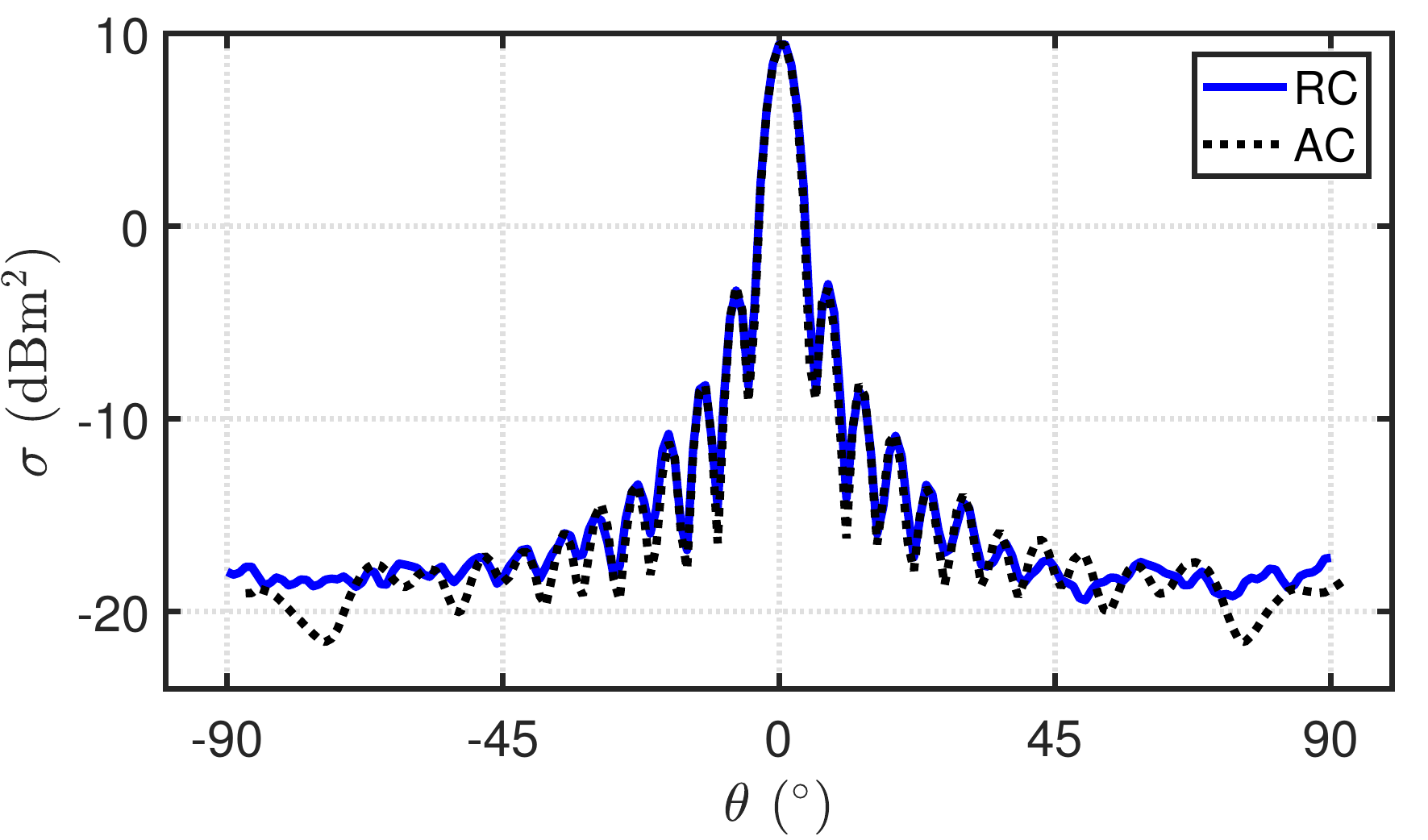}}
\caption{\pb{Comparison of RCS extracted from RC measurements using a distance bandwidth of $4$~GHz between $8$~GHz and $12$~GHz and an  amplitude bandwidth of $0.9$~GHz for different central frequencies ($9$~GHz and $11$~GHz) and RCS obtained from AC measurements according to the plate  position in the $-90^\circ$ to $+90^\circ$ azimuth range  and in the VV polarization.}}
\label{difference_f_VV}
\end{figure}

The measurement data in the VV configuration are then processed to extract the RCS pattern at $9$~GHz and $11$~GHz. The obtained results are shown in the Fig.~\ref{difference_f_VV}. \fs{Again, RCS local minima and maxima are well retrieved and a satisfying agreement with the AC measurement is observed.}

\section{Conclusion}

This paper deals with the post-processing enhancement of an RCS measurement method performed within RCs. The method is based on a differential measurement (with and without the target) in a given frequency band, on which a sinusoidal regression is applied in order to extract the RCS amplitude. A first step consists in evaluating the antenna-target distance which sets the sine period, whereas a second step is dedicated to the RCS amplitude estimation. While the frequency bandwidth has been arbitrarily chosen in previous works, this article investigates the proper selection of the bandwidth used for distance and RCS amplitude estimations.

It is shown that a large frequency bandwidth allows a better estimation of the antenna-target distance, by providing a finer distance resolution. In particular, this allows discriminating multiple distinct scattering zones, as for example the two edges of a plate. However, it is also shown that enlarging the frequency bandwidth leads to a frequency-averaged RCS estimation. Indeed, although it does not affect much the main lobe retrieval, it strongly degrades the rest of the RCS pattern by smoothing out the secondary lobes. Therefore, a two-step method is introduced in order to take into account these two contradictory constraints.

The two-step method consists in defining two distinct frequency bandwidths: one for the distance estimation (the distance bandwidth) and one for the amplitude estimation (the amplitude bandwidth). The distance estimation is firstly performed using an as-large-as-possible distance bandwidth in order to better estimate the position of the main scattering contribution. Once the distance has been precisely evaluated, the RCS amplitude estimation is secondly performed using a reduced amplitude bandwidth in order to avoid the frequency-averaging effect. However, although an as-narrow-as-possible amplitude bandwidth is theoretically targeted, a minimum bandwidth is mandatory in order to ensure a sufficient number of RC uncorrelated states (on which the regression accuracy relies). It has been shown that in the case of our specific setup, a $0.9$~GHz amplitude bandwidth exhibits the best trade-off, while the distance bandwidth is fixed to 4~GHz.

\fs{RCS patterns of a rectangular metallic plate (152~ mm $\times$ 150~mm) at $10$~GHz measured in an RC for $-90^\circ \leq \theta \leq 90^\circ$ (VV and HH polarizations) have been compared to the ones measured in an AC. Compared to the one-step method, the newly introduced two-step method enables reducing the average relative difference from $234\%$ to $50\%$ in the HH polarization and from $32\%$ to $20\%$ in the VV polarization.} These results allow considering RCs as a reliable alternative to anechoic chambers for RCS measurements. \pb{The reliability and accuracy of the proposed method is confirmed for targets assimilated as a point-like scatterer.  Future works will be dedicated to the study of resonating or multiple point-like scatterers.}




 \section*{Acknowledgment}
This work is supported by AID/DGA, France. It is also supported in part by the European Union through the European Regional Development Fund, in part by the Ministry of Higher Education and Research, in part by the Région Bretagne, and in part by the Département d'Ille et Vilaine through the CPER Project SOPHIE/STIC \& Ondes.

\end{document}